\pdfoutput=1
\documentclass[
reprint,superscriptaddress,amsmath,amssymb,aps,prd,floatfix,footinbib]{revtex4-2}
\usepackage[a4paper,left=1.5cm,right=1.5cm,top=3cm,bottom=3cm]{geometry}
\usepackage[toc]{appendix} 

\usepackage{graphicx}
\usepackage{dcolumn}
\usepackage{bm}
\usepackage{amssymb,amsmath,amsfonts,physics,epsfig}
\usepackage[dvipsnames]{xcolor}
\usepackage{longtable}
\usepackage{verbatim}
\usepackage{color}
\usepackage{mdframed}
\usepackage{psfrag}
\usepackage{soul}
\usepackage{bm}
\usepackage{amsfonts,amssymb,mathrsfs,amsmath,esint}
\usepackage{slashed, cancel}
\usepackage{framed}
\usepackage{mdframed}
\usepackage{physics}
\usepackage{simplewick} 
\allowdisplaybreaks
\usepackage{latexsym}
\graphicspath{{./Figures/}}
\usepackage[dvipsnames]{xcolor}
\usepackage{booktabs}
\usepackage{datetime}
\newdateformat{mydate}{\THEDAY{ }\monthname[\THEMONTH]{ }\THEYEAR}

\usepackage{tikz}
\usepackage{color}
\usepackage{framed}
\usepackage{hyperref}
\usepackage[capitalise]{cleveref}
\hypersetup{colorlinks, citecolor=bluscuro, linkcolor=black, urlcolor=bluscuro}
\definecolor{rossos}{cmyk}{0,1,1,0.55}
\definecolor{bluscuro}{rgb}{0.15, 0.2, .85}
\definecolor{bluchiaro}{cmyk}{1,.3,0.,0.1}

\graphicspath{{./images/}}

\newcommand{\be}{\begin{equation}}
	\newcommand{\ee}{\end{equation}}
\newcommand{\bea}{\begin{eqnarray}}
	\newcommand{\eea}{\end{eqnarray}}
\newcommand{\beq}{\begin{equation}}
	\newcommand{\eeq}{\end{equation}}

\def\beqa{\begin{eqnarray}}

	\def\eeqa{\end{eqnarray}}

\def\lsim{\mathrel{\rlap{\lower4pt\hbox{\hskip0.5pt$\sim$}}
		\raise1pt\hbox{$<$}}}         
\def\gsim{\mathrel{\rlap{\lower4pt\hbox{\hskip0.5pt$\sim$}}
		\raise1pt\hbox{$>$}}}         

\makeatletter
\newcommand*{\rom}[1]{\expandafter\@slowromancap\romannumeral #1@}
\newcommand{\Den}{
  \left[ -1 + G H(t)^{2} \right]
  \left[ -1 + G H(t)^{2} + G \dot{H}(t) \right]^{2}
  \left[ -2 + 3 G H(t)^{2} + G \dot{H}(t) \right]^{2}
  \left[ -2 + 3 G H(t)^{2} + 3 G \dot{H}(t) \right]^{2}
}
\makeatother

\usepackage[normalem]{ulem}
\usepackage{soul}

\begin{document}
	\title{Inflation from entropy}
	
	\author{Udaykrishna Thattarampilly}
	\email{uday7adat@gmail.com}
    
	\author{Yunlong Zheng}
	\email{zhyunl@yzu.edu.cn (corresponding author)}

    	\affiliation{Center for Gravitation and Cosmology, College of Physical Science and Technology, Yangzhou University, Yangzhou 225009, China}
	\date{\today}

	\begin{abstract}
		We investigate cosmological solutions for the modified gravity theory obtained from quantum relative entropy between the metric of spacetime and the metric induced by the geometry and matter fields. The vacuum equations admit inflationary solutions, hinting at an entropic origin for inflation. Equations also admit a regime of phantom like behavior. Assuming that the relation between slow roll parameters and CMB observables holds for “gravity from entropy”, the theory predicts a viable spectrum. 
  \end{abstract}
	
	\maketitle
\section{Introduction}	
\label{se:intro}
	 Quantum information and its relation to gravity have been of key interest to theoretical physicists since the discovery of black hole entropy \cite{Bekenstein:2008smd,Bekenstein:1974ax} and Hawking radiation \cite{Hawking:1975vcx}. The discovery of black hole entropy implied that gravity inherently encodes quantum information, since entropy quantifies the number of microscopic states hidden from an observer. This connection has deepened with the discovery of the holographic principle \cite{tHooft:1999rgb,Susskind:1994vu,Swingle:2009bg}, and recent advances in entanglement entropy \cite{Ryu:2006ef,Nishioka:2009un,Faulkner:2013ana,Witten:2018zxz,Sorce:2023fdx,Ben-Dayan:2023inz}. These developments challenge classical notions of gravity as purely geometric, instead positing that gravitational forces may emerge from the collective entanglement entropy of quantum degrees of freedom, redefining gravity as a thermodynamic or information-theoretic phenomenon \cite{Padmanabhan:2009vy}.  

A comprehensive information-theoretic approach to gravity is expected to provide deeper insights into the early Universe Cosmology, Black hole physics \cite{Barack:2018yly}, and quantum gravity \cite{PhysRevLett.93.131301,Oriti_2009}. A recent study has proposed that the quantum relative entropy between the spacetime metric and the metric induced by geometry and matter fields serves as the fundamental action governing the theory of gravity \cite{Bianconi:2024aju}. Quantum relative entropy \cite{Vedral:2002zz} is a key concept in information theory and is defined for quantum operators \cite{Araki:1976zv,10.1093/oso/9780198517733.002.0001,ohya1993quantum}. In the proposed theory \cite{Bianconi:2024aju}, the spacetime metric, the geometry-induced metric, and matter fields are treated as quantum operators, forming a bimetric theory of gravitation \cite{Rosen1973ABT,Hossenfelder:2008bg} where metrics are promoted to quantum operators. To distinguish this framework from other formulations of entropic gravity, we refer to it throughout this paper as the “Gravity from Entropy” approach.
The gravity from entropy approach leads to modified Einstein equations that reduce to classical general relativity in the weak-coupling and low-curvature limits \cite{Bianconi:2024aju}. 

Recent work on approximate Schwarzschild solutions in entropic quantum gravity concluded that black holes with lengths far exceeding the Planck length obey the area law for entropy \cite{Bianconi:2025rnd}. In this article, we derive the equivalent of the Friedmann equations for gravity from entropy. The equations admit inflationary solutions in the absence of additional matter fields. Inflation is the brief epoch of exponential expansion in the universe's earliest moments \cite{Guth:1980zm}, resolving the horizon and flatness problems while seeding primordial density fluctuations \cite{Linde:1981mu,Linde:2007fr,Martin:2013tda,Baumann:2014nda,Guth:1982ec,Kodama:1984ziu,Baumann:2009ds,Cheung:2007st}. These fluctuations are imprinted on the CMB as temperature fluctuations. The nearly scale-invariant spectrum of these fluctuations is largely in agreement with standard inflationary predictions.

Standard inflation models are driven by a hypothetical inflaton field\cite{Linde:1983gd}. The only scalar field observed so far is the Higgs field \cite{Bezrukov:2007ep}. A minimal coupling to the Higgs field does not admit standard slow-roll solutions \cite{Linde:1983gd}, and nonminimal coupling \cite{Cervantes-Cota:1995ehs} leads to non-renormalizable corrections \cite{Burgess:2009ea,Burgess:2010zq} and unitarity violation. An alternative way to achieve inflation is modifications of Gravity, often with higher order curvature corrections \cite{Nojiri:2017ncd,Capozziello:2011et,Nojiri:2010wj,Olmo:2011uz}. All standard inflation models suffer from fine-tuning and the inability to prescribe a unique measure \cite{Guth:1980zm,Ijjas:2013vea}. Although inflation starts out high in its potential, there doesn't exist a theory of initial conditions to explain it. This issue is related to the low entropy initial state required for inflation, known as the entropy problem \cite{Linde:2014nna,Ijjas:2013vea}. Although highly successful, the theoretical challenges involved reconciling inflation with UV theories and the lack of predictability have prompted physicists to search for alternative models of the early Universe, such as bounce \cite{Battefeld:2014uga,Novello:2008ra,Ben-Dayan:2023rlj,Brandenberger:2009jq,Lehners:2008vx,Ijjas:2019pyf,Artymowski:2020pci}.  

FLRW solutions to entropic quantum gravity are naturally inflationary without additional terms or matter fields. Gravity emerges from the Von Neumann entropy of quantum operators, endowing the model with UV completion and robust theoretical motivation. Inflation in entropic quantum gravity does not involve any additional parameters apart from Newton's constant $G$. As we shall observe later in the paper, inflation can occur as both low and high-entropy solutions for the equations of motion.
High entropy solutions correspond to a Hubble parameter (H) of $ \sqrt{0.12} M_{pl} \lsim H < \sqrt{1/6} M_{pl}$  and predict a tensor to scalar ratio $0.010 \lsim r \lsim 0.012 $ with the spectral index $0.962 \pm 0.002$ for CMB observables. Solutions exhibit a phantom behavior for $\sqrt{0.08} M_{pl} \lsim H \lsim 0.11 M_{pl}$. The slow roll parameter $\epsilon$ is small and negative in this regime.

This paper is organized as follows: first, we discuss gravity from entropy and equations of motion for the FLRW-like Universe in the absence of matter fields. In the following sections, we solve the equations of motion, demonstrating that spacetime is inflationary. This is followed by a section on numerical solutions and the regime of phantom behavior.\\ 

\section{Gravity from entropy}
\label{sec:ge}
The gravity from entropy theory proposed in \cite{Bianconi:2024aju} involves a topological metric composed of metrics between scalars, vectors, and bi-vectors defined on a 4-d manifold fully described by the metric $g_{\mu \nu}$. The form of the topological metric is 
\begin{equation}
    \tilde{g} = 1 \oplus g_{\mu \nu} dx^{\mu} \otimes dx^{\nu} \oplus g^{(2)}_{\mu \nu \rho \sigma} (dx^{\mu} \wedge dx^{\nu}) \otimes (dx^{\rho} \wedge dx^{\sigma}) 
\end{equation}
where $ g^{(2)}_{\mu \nu \rho \sigma} =\frac{1}{2} \left(g_{\mu \rho} g_{\nu\sigma}-g_{\mu\sigma} g_{\nu\rho} \right)$. An additional metric induced by the geometry and matter fields $\tilde{\mathbf{G}}$ is introduced as a direct sum of a metric between scalars $\tilde{G}^{(0)}$, a metric between vectors $\tilde{G}^{(1)}_{\mu \nu}$, and a metric between bi vectors $\tilde{G}^{(2)}_{\mu \nu \rho \sigma}$. 
\begin{equation}
    \tilde{\mathbf{G}} = \tilde{G}^{(0)} \oplus \tilde{G}^{(1)}_{\mu \nu} dx^{\mu} dx^{\nu} \oplus \tilde{G}^{(2)}_{\mu \nu \rho \sigma} (dx^{\mu} \wedge dx^{\nu}) \otimes (dx^{\rho} \wedge dx^{\sigma}) 
\end{equation}
where $\tilde{G}^{(0)}$, $\tilde{G}^{(1)}$ and $\tilde{G}^{(2)}$ are invertible at every point on the manifold. 

The approach of quantum relative entropy proposes a modified gravity action given by 
\begin{equation}
    S = \frac{1}{\left(\textit{l}_{pl}\right)^4} \int \sqrt{-g} \mathcal{L} d^4 x 
    \label{eq:action}
\end{equation}
where $\textit{l}_{pl} = \left(\frac{\hbar G}{c^3} \right)^{1/2}$ is the plank length and $\mathcal{L}$, the Lagrangian density is 
\begin{equation}
    \mathcal{L} = -\mathrm{Tr} \log\left( \tilde{\mathbf{G}} \tilde{g}^{-1} \right).
    \label{eq:lagrangian}
\end{equation}
The entropic action can be expressed in terms of the modular operator $\Delta^{\frac{1}{2}}_{\tilde{\mathbf{G},\tilde{g}}}$, since
\begin{equation}
  \tilde{\mathbf{G}} \tilde{g}^{-1} = \Delta^{\frac{1}{2}}_{\tilde{\mathbf{G},\tilde{g}}} = \sqrt{\tilde{\mathbf{G}} \tilde{\mathbf{G}}\star}
\end{equation}
and is a generalization of the Araki quantum relative entropy \cite{Araki:1976zv}. The metrics $\tilde{g}$ and $\tilde{\mathbf{G}}$ are treated as ``renormalizable" density matrices at every point on the manifold, and the action is the relative entropy between them (Refer \cite{Bianconi:2024aju, Bianconi:2024hts} for details). Variation of the action with respect to the spacetime metric leads to modified Einstein equations that reduce to Einstein gravity in the low energy limit \cite{Bianconi:2024aju}.  

The metric induced by the geometry in the vacuum is assumed to be \cite{Bianconi:2024aju,Bianconi:2025rnd} 
\begin{equation}
     \tilde{\mathbf{G}} = \tilde{g}-\frac{G}{2} \tilde{\mathbf{\mathcal{R}}} 
     \label{eq:gfie}
\end{equation}
where G is Newton's constant (we have chosen the coupling $\alpha = \frac{G}{2}$, $\beta = \frac{1}{2}$ according to the notation used in the paper \cite{Bianconi:2025rnd}, see Appendix \ref{sec:appendix1} for details. This coupling was originally chosen to match the coupling in an earlier version of the article \cite{Bianconi:2025rnd}. An alternative choice of coupling will result in inflation occurring at a different value of the Hubble parameter, as explained later in the paper) and 
\begin{equation}
     \tilde{\mathbf{\mathcal{R}}} = R \oplus (R_{\mu \nu}dx^{\mu }\otimes dx^{\nu} )\oplus R_{\mu\nu\rho\sigma} (dx^{\mu} \wedge dx^{\nu}) \otimes (dx^{\rho} \wedge dx^{\sigma}).
\end{equation}
$R$ is the Ricci scalar, $R_{\mu\nu}$ is the Ricci tensor, and $g^{\mu \eta} R_{\eta\nu\rho\sigma} $ is the Riemann tensor. The general form of modified vacuum Einstein equations is obtained in \cite{Bianconi:2024hts}. If we restrict ourselves to the special class of diagonal metrics for which $R_{\mu\nu\rho\sigma} \neq 0$ if $\mu \neq \nu$, $\mu=\rho$, $\nu=\sigma$, or if $\mu=\sigma$ and $\nu=\rho$ the modified vacuum Einstein equations obtained by variation of the action are simpler and can be solved with some approximations and assumptions. Important metric spacetimes in physics, such as FLRW and Schwarzschild, fall into this category. In this work, we derive and solve the modified field equations for the FLRW spacetime in a vacuum and in the presence of a real scalar field. We rigorously demonstrate that the resulting solutions inherently exhibit exponential expansion without introducing ad hoc corrections, exotic matter fields, or supplementary inflationary mechanisms. This outcome arises purely from the geometric structure of the theory. \\

\section{Modified vacuum Einstein equations}  
The product of the metric induced by the geometry of spacetime and the inverse topological metric $\tilde{g}^{-1}$ is
\begin{equation}
\begin{split}
   \tilde{\mathbf{G}} & \tilde{g}^{-1} = \mathbf{I}-\frac{G}{2} \tilde{\mathbf{\mathcal{R}}}  \tilde{g}^{-1} = (1-\frac{G}{2}R) \oplus \left( \delta_{\mu}^{\nu} -\frac{G}{2}R_\mu^{\nu} \right)\; dx^{\mu} \otimes dx_{\nu} \\
    &\;\;\oplus \left(\frac{1}{2}
    \delta_{\mu\nu}^{\rho\sigma} - \frac{G}{2} R_{\mu\nu}^{\;\;\;\rho \sigma}\right)  (dx^{\mu} \wedge dx^{\nu}) \otimes (dx_{\rho} \wedge dx_{\sigma}) = \mathbf{\mathcal{G}}^{-1}
    \end{split}
\end{equation}
The Lagrangian for gravity from entropy described in equation \eqref{eq:lagrangian} involves the trace of a tensor $\tilde{G}^{(2)}_{\mu \nu \rho \sigma} g^{(2)\rho \sigma \eta \theta}$. The trace is defined as the trace of the flattened $6 \times 6$ matrix $\tilde{G}^{(2F)}_{\mu \nu \rho \sigma}g^{(2F)\rho \sigma \eta \theta}$ where the superscript $F$ denotes the flattened matrix obtained from a tensor (see Appendix B of article \cite{Bianconi:2024aju} for details).

In the absence of matter fields $\tilde{G}^{(2F)}_{\mu \nu \eta \theta}  g^{(2F)\eta \theta \rho \sigma }= 
    \delta_{\mu\nu}^{\rho\sigma}- \frac{G}{2} R_{\mu\nu}^{\;\;\;\rho \sigma}$. By demanding that the entries of the Riemann curvature are non-zero only when $\mu \neq \nu$, $\mu=\rho$ and $\nu=\sigma$ (and the permutation $\mu=\sigma$ and $\nu=\rho$ ), we ensure that the corresponding flattened matrix is diagonal. Only non-zero entries of $R_{\mu\nu}^{\;\;\;\rho \sigma}$ are of the form
$R_{\mu\nu}^{\;\;\;\mu \nu}$, $\mu \neq \nu$  (the indices here are not summed over). Consequently, the Ricci tensor $R_{\mu\nu}$ is also diagonal. The Lagrangian in \eqref{eq:lagrangian} can then be expressed as 
\begin{equation}
\begin{split}
    -\mathrm{Tr} &\log\left( \mathbf{I}-\frac{G}{2} \tilde{\mathbf{\mathcal{R}}} \tilde{g}^{-1}   \right) = -\log\left(1- \frac{G}{2} R\right)-\\&\sum_{\mu\nu}^{\mu=\nu}\log\left( \delta_{\mu}^{\nu} -\frac{G}{2}R_\mu^{\nu}\right)-\sum^{\mu<\nu}_{\mu \nu} \log \left(
    \delta_{\mu\nu}^{\mu\nu}- G R_{\mu\nu}^{\;\;\;\mu \nu} \right) .
    \end{split}
      \label{eq:logr}
\end{equation}
The expression in equation \eqref{eq:logr} is valid since the flattened matrix and the Ricci tensor are both already diagonal. \\
Since both the flattened matrix and the Ricci tensor are diagonal, we can write
\begin{equation}
\begin{split}
   \mathbf{\mathcal{G}} &  = \frac{1}{(1-\frac{G}{2}R)} \oplus \frac{1}{\left( \delta_{\mu}^{\nu} -\frac{G}{2}R_\mu^{\nu} \right)}\; dx^{\mu} \otimes dx_{\nu} \\
    &\;\;\oplus \frac{1}{\left(\frac{1}{2}
    \delta_{\mu\nu}^{\rho\sigma} - \frac{G}{2} R_{\mu\nu}^{\;\;\;\rho \sigma}\right)}  (dx^{\mu} \wedge dx^{\nu}) \otimes (dx_{\rho} \wedge dx_{\sigma}) 
    \end{split}
\end{equation}
where $ \mathbf{\mathcal{G}}$ is defined as
\begin{equation}
     \mathbf{\mathcal{G}}^{-1} = \mathbf{I}-\frac{G}{2} \tilde{\mathbf{\mathcal{R}}} 
\end{equation}
The modified vacuum Einstein equation for the theory is given by \cite{Bianconi:2024aju}
\begin{equation}
    R^{\mathcal{G}}_{\mu\nu}-\frac{1}{2}  g_{\mu\nu}\left(R_{\mathcal{G}}-2\Lambda_{\mathcal{G}}\right)+D_{\mu\nu}=0
    \label{eq:modein}
\end{equation}
where
\begin{equation}
    R_{\mathcal{G}} =  -\mathrm{Tr} \left(g^{-1}\mathcal{G} \tilde{\mathcal{R}}\right),
\end{equation}
\begin{equation}
    \Lambda_{\mathcal{G}} =  -\frac{1}{2G}\mathrm{Tr} \left(\tilde{\mathcal{G}}- \tilde{\mathcal{I}}-\log\left(\tilde{\mathcal{G}}\right)\right),
\end{equation}
\begin{equation}
\begin{split}
R^{\mathcal{G}}_{\mu\nu} 
&= \mathcal{G}_{(0)}\, R_{\mu\nu} 
+ \left[ \mathcal{G}_{(1)} \right]^{\rho}{}_{\mu} R_{\rho\nu} \\
&\quad - \left[ \mathcal{G}_{(2)} \right]_{\rho_1 \rho_2 \mu \nu} R^{\rho_1 \rho_2} 
+ 2\, \left[ \mathcal{G}_{(2)} \right]^{\eta \rho_1 \rho_2}{}_{\mu} R_{\rho_1 \rho_2 \nu \eta}
\end{split}
\end{equation}
and
\begin{equation}
\begin{split}
D_{\mu \nu} &= \left( \nabla^{\rho} \nabla_{\rho} g_{\mu \nu}
- \nabla_{\mu} \nabla_{\nu} \right) \mathcal{G}_{(0)}
- \nabla^{\rho} \nabla_{\nu} \left[ \mathcal{G}_{(1)} \right]_{(\rho \mu)} \\
&\quad + \frac{1}{2} \nabla^{\rho} \nabla_{\rho} 
\left[ \mathcal{G}_{(1)} \right]_{\mu \nu}
+ \frac{1}{2} \nabla^{\eta} \nabla_{\eta} 
\left[ \mathcal{G}_{(1)} \right]_{\rho \nu} g_{\mu \nu} \\
&\quad + \nabla^{\eta} \nabla^{\nu} 
\left[ \mathcal{G}_{(2)} \right]_{\mu \rho \eta}
+ \nabla^{\eta} \nabla^{\nu} 
\left[ \mathcal{G}_{(2)} \right]_{\eta \mu \nu} \\
&\quad + \frac{1}{2} \left[ \nabla^{\rho}, \nabla^{\eta} \right] 
\left[ \mathcal{G}_{(2)} \right]_{\rho \eta \mu \nu}
\end{split}
\end{equation}
\section{Modified Friedmann equations} Assuming that the spacetime is FLRW, equation \eqref{eq:modein} reduces to a pair of coupled ordinary differential equations dubbed modified Friedmann equations. The modified Friedmann equations for entropic gravity are highly nonlinear and are given by
\begin{equation}
    \begin{split}
    R^{\mathcal{G}}_{00}+\frac{1}{2}  \left(R_{\mathcal{G}}-2\Lambda_{\mathcal{G}}\right)+D_{00}=0\\
     \frac{1}{g_{11}} R^{\mathcal{G}}_{11}+\frac{1}{2}  \left(R_{\mathcal{G}}-2\Lambda_{\mathcal{G}}\right)+\frac{D_{11}}{g_{11}}=0.
      \label{eq:modfried}
    \end{split}
\end{equation}
We refer the reader to the Appendix \ref{sec:app2}, for expanded expressions of all the terms in the Friedmann equations. Equations concerning $ R^{\mathcal{G}}_{22}$ and $ R^{\mathcal{G}}_{33}$ are the same as the second equation. 
\\

\section{Inflation from entropy}
The modified Friedmann equations for entropic gravity admit inflationary solutions. Since inflation drives the spatial curvature near zero within a few e-folds, we mostly concern ourselves with a spatially flat Universe. In Appendix \ref{sec:app2a} we rewrite the modified Friedmann equations for a flat Universe in terms of the Hubble parameter. 
The equations are reduced to the usual Friedmann equations when the higher-order terms are negligible. 

From the time component of equation \eqref{eq:modfried} we obtain an equation for $H''(t)$ and by substituting this equation in the second Friedmann equation we obtain the equation of motion for the FLRW Universe in entropic gravity. 


We rewrite equation \eqref{eq:modfried} in terms of slow roll parameters $\epsilon$ defined as
\begin{equation}
    \epsilon = -\frac{\dot{H}}{H^2} = -\frac{H'(N)}{H(N)} .
\end{equation}
where $N$ is the number of e-folds of inflation. Assuming a slow roll scenario where $H=e^{-N \epsilon}$ and $\epsilon \ll 1$, we expand the equations to the first order in $\epsilon$ to determine $\epsilon$ and $\eta$ as functions of $N$ and $x=G H^2$. In general, when $\beta$ is a free parameter $x=2\beta G H^2$, and the value of Hubble corresponding to inflation will vary accordingly. The results discussed here for $\beta=\frac{1}{2}$ hold invariably for all values of $\beta$ with a corresponding modification in the value of $H$. Our analysis showed that the solutions of the equations for entropic gravity can be divided into two different regimes corresponding to $ 0 < x \lsim 0.08$ and $ 0.08 \lsim x \lsim 1/6$. The equations admit inflationary solutions for $G H^2 < 0.08 $ and for $ 0.12 \lsim x < 1/6$. Near $x \sim 0.08$ $\epsilon$, the drop is large, and the slow roll assumption is broken.

The regime of $x \ll 1/6$ corresponds to $\mathcal{G}\sim \mathbf{I}$ where the Cosmological constant $\Lambda_{\mathcal{G}}$ is small \cite{Bianconi:2024aju}. Accelerated expansion is expected for $x \ll 1/6$ due to the presence of this small cosmological constant. 
 $\Lambda_{\mathcal{G}}$ is divergent for $\mathcal{G} \sim 0$ and for $\Lambda_{\mathcal{G}}$ approaching infinity. This is the relevant regime to study for large Hubble (such as in the early Universe). As $x$ approach a value of $\frac{1}{6}$, $\mathcal{G} \sim 0$, and $\Lambda_{\mathcal{G}}$ becomes large and significant.
The correction terms coming from gravity from entropy theory are divergent as $x$ approaches $1/6$, leading to a scenario similar to inflation where the extremely large potential drives the accelerated expansion. The inflationary solution near $x \sim 1/6$ is rather unexpected and, more importantly, is consistent with inflationary observables. 

\subsection{Slow roll inflation for $x \ll 1/6$}
Assuming a slow roll scenario with nearly constant $\epsilon$ and $H$,
when $x\sim0$, we have
\begin{equation}
\epsilon \sim \frac{3}{2 \left( -8 + 3N \right)}
- \frac{(11 + 3N)\,x}{4 \left( -8 + 3N \right)^{2}}.
\label{eq:epn}
\end{equation}
See Appendix \ref{sec:app3} for details and the full expression for $\epsilon$.
The second slow roll parameter $\eta$ is defined as
\begin{equation}
    \eta = -\dfrac{d\epsilon}{dN}.
\end{equation}
For $x\ll1$ and $N\gg0$, $\epsilon\ll1$ and $\eta\ll1$, indicating that we have slow roll inflation without introducing additional matter fields.  

\begin{figure}[!ht]
    \centering
    \includegraphics[scale=0.53]{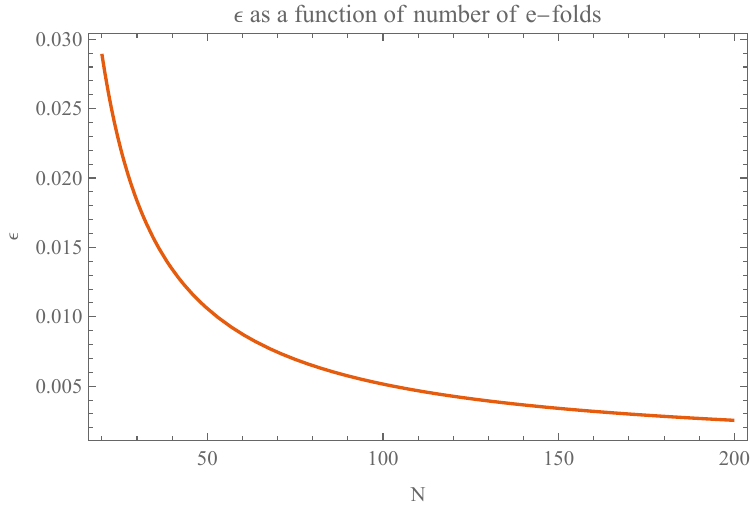}
    \includegraphics[scale=0.53]{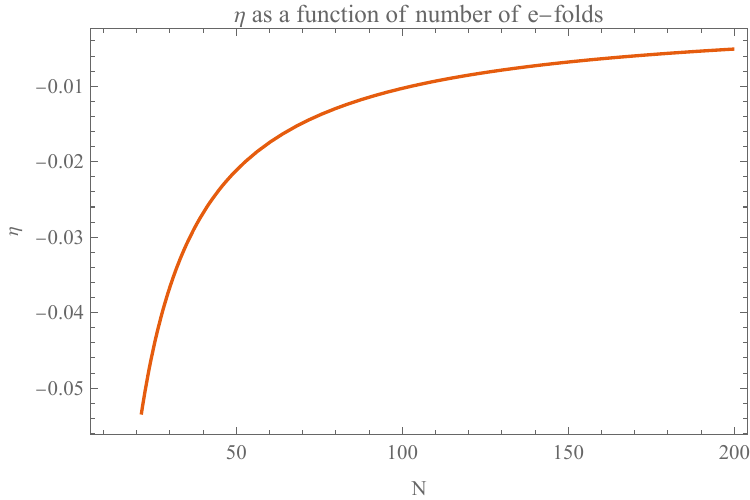}
    \caption{ $\epsilon$ and $\eta$ as a function of number of e-folds for $x=10^{-2}$. Solutions represent a slow roll scenario where $\abs{\eta} \ll1$ and $\epsilon \ll1$}.
    \label{fig:usr}.
\end{figure}
Figure (\ref{fig:usr}) depicts $\epsilon$ and $\eta$ as a function of the number of e-folds. The solutions are inflationary and well within the slowroll regime. Assuming that the relation between slowroll parameters and CMB observables $r$ and $n_s$ remains intact for entropic gravity, $x\sim 0$ solutions are invalidated by observations. The observed bound on the tensor to scalar ratio is $r\lsim 0.036$ and the spectral index is $n_s \sim 0.96 \pm 0.0042$ \cite{Planck:2018vyg,Planck:2019kim,Planck:2018jri}.

\subsubsection{Numerical solutions} 
We solve the equations \eqref{eq:modfried} numerically for $x\ll 1$. 
\begin{figure}[!ht]
    \centering
    \includegraphics[scale=0.53]{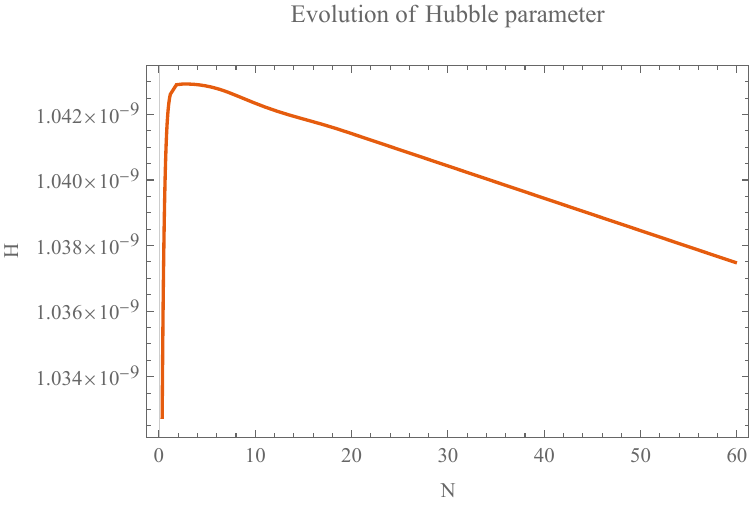}
    \includegraphics[scale=0.53]{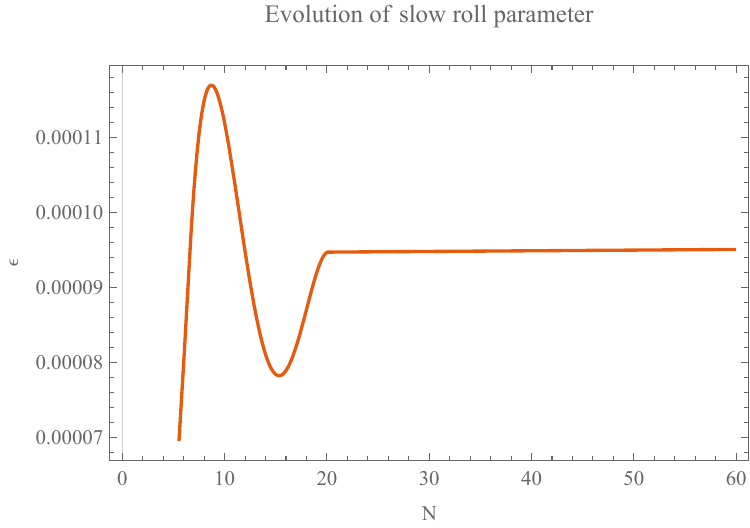}
     \caption{Evolution of the Hubble (top) and slowroll (bottom) parameters with number of e-folds N starting from for initial conditions $H(0)=10^{-9}$ and $H'(0)=3/16 \times 10^{-9}$}.
    \label{fig:sr epsilom}
\end{figure}
In figure (\ref{fig:sr epsilom}) we have numerical solutions for the modified Friedmann equations with initial conditions $H(0)=10^{-9}$ and $H'(0)=3/16 \times 10^{-9}$. The second Friedmann equation sets the initial condition for the derivative. As observed in the picture, the Hubble parameter is increasing initially, but quickly approaches a constant with a slight decrease over time. The parameter $\epsilon$ is positive, small, and is a constant after a few e-folds. The numerical analysis confirms our analytical results regarding slow roll inflation. 

\subsection{A viable solution: Slow roll inflation for $ \textcolor{red}{x} \sim 1/6$}
For larger values of $x$ the approximation in equation \eqref{eq:epn} is not valid. However, from the full expression for $\epsilon$ expressed in Appendix \rom{3} we observe that the equation allows inflationary solutions near $x=1/6$. 
\begin{figure}[!ht]
    \centering
    \includegraphics[scale=0.53]{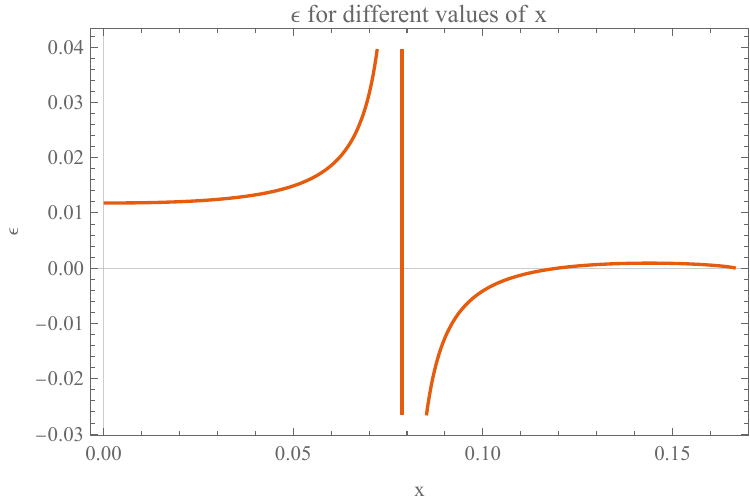}
    \includegraphics[scale=0.53]{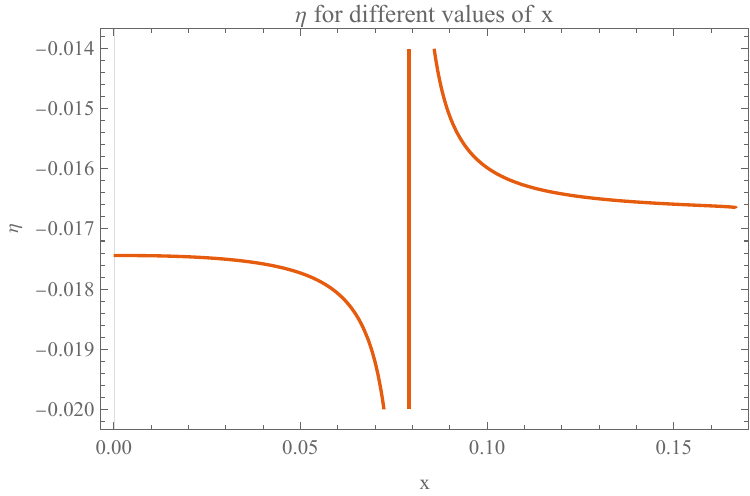}
    \caption{ $\epsilon$ and $\eta$ as a function  \textcolor{red}{$x$}  after 50 e folds of inflation. Solutions are slow rolling with $\abs{\eta} \ll1$ and $0<\epsilon \ll1$ for $0<x\lsim0.08$ and $0.12 \lsim x <1/6$.}
    \label{fig:srl}.
\end{figure}
Figure (\ref{fig:srl}) shows the slow roll parameters $\epsilon$ and $\eta$ for different values of $0 < x < 1/6$ after 50 e-folds of inflation. It is clear from the figure that the solution exhibits a phantom like behavior for $0.08 < x \lsim 0.12$ and is inflationary for $0.12 \lsim x < 1/6$. For $x > 1/6$ the action is complex, and the equations for gravity from entropy are no longer valid. 

Incredibly, the inflationary solutions for $0.12 \lsim x < 1/6$ produce a tensor to scalar ratio well within the observed bound of $r\lsim 0.036$ and have a spectral index of $0.96 \lsim n_s \lsim 0.964$ in agreement with observations. Since the Universe is undergoing a slow roll inflation, we have assumed that the standard results of $r=16\epsilon$ and $n_s=1-4\epsilon+2\eta$ \cite{Kinney:2003xf,Baumann:2009ds,Liddle_Lyth_2000}, hold for gravity from entropy theory.

\begin{figure}[!ht]
    \centering
    \includegraphics[scale=0.53]{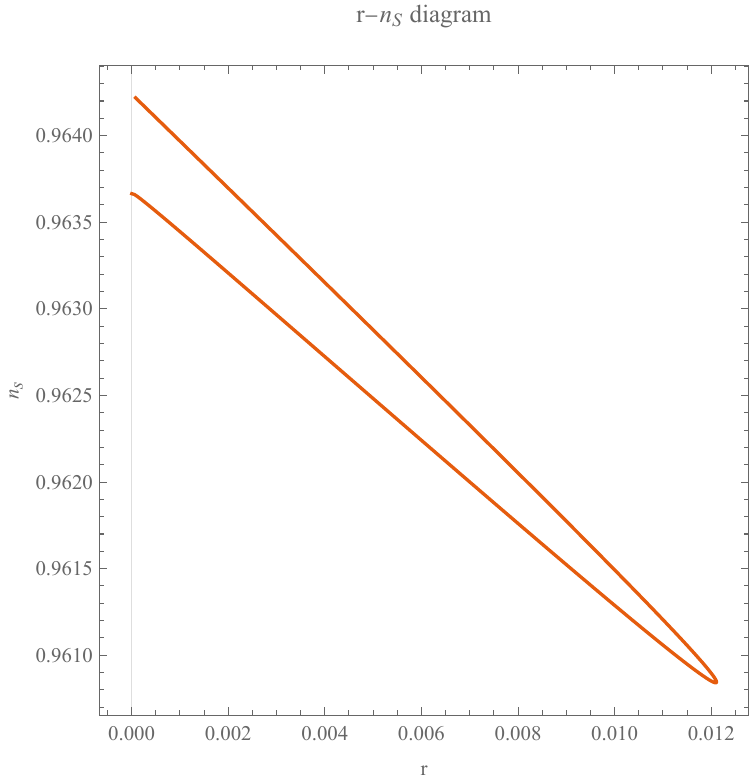}
    \caption{Tensor to scalar ratio vs spectral index plot for $0.12 \lsim x < 1/6$ after 55 e-folds of inflation.}
    \label{fig:nsr}.
\end{figure}
 Figure (\ref{fig:nsr}) is a tensor to scalar ratio vs spectral index plot for $0.12 \lsim x < 1/6$ assuming 55 e-folds of inflation. $n_s$ is in agreement with the CMB observations and the tensor to scalar ratio $0.000142 \lsim r \lsim 0. 012$ potentially testable by experiments such as SO or CMB-S4. The lower limit on $r$ depends on the number of e-folds of inflation, for 60 e-folds of inflation $0.010 \lsim r \lsim 0.012$, values of $n_s$ that fall within the observed range.  
 
The leading terms of the expressions for $\epsilon$ and $\eta$ near $x=1/6$ are
 \begin{equation}
\epsilon \sim \frac{
    1.5 \,
    \bigl( 2.24 + \ln\!\left[ \frac{\tfrac{1}{6}}{\tfrac{1}{6} - x} \right] \bigr)
    \bigl( -1.98 + \ln\!\left[ \frac{\tfrac{1}{6}}{\tfrac{1}{6} -x} \right] \bigr)
    \left(\tfrac{1}{6}-x \right)
}{
    0.55
    + 0.13\,N
    + N \, \ln\!\left[ \frac{\tfrac{1}{6}}{\tfrac{1}{6} - x} \right]
}
\end{equation}
and
\begin{equation}
\eta \sim -\frac{
    0.13 + \ln\!\left[ \frac{\tfrac{1}{6}}{x-\tfrac{1}{6}} \right]
}{
    0.55
    + 0.13\,N
    + N \, \ln\!\left[ \frac{\tfrac{1}{6}}{x-\tfrac{1}{6} } \right]
}.
\end{equation}
 The results discussed here regarding CMB observables is based on background dynamics. A comprehensive study of perturbations in entropic gravity is yet to be realized. Further analysis based on such a theory is required to determine the value of observables conclusively. This is beyond the scope of this work and is left for future endeavors.\\

\section{Phantom like behavior from entropy}
For $ 0.08 \lsim x \lsim 0.11$, the slow roll parameter $\epsilon$ is negative, indicating an equation of state $w<-1$, similar to phantom dark energy scenarios \cite{Caldwell:2003vq}, or in some modified gravity theories \cite{Kofinas:2014daa,Granda:2011kx}. 
\begin{figure}[!ht]
    \centering
    \includegraphics[scale=0.53]{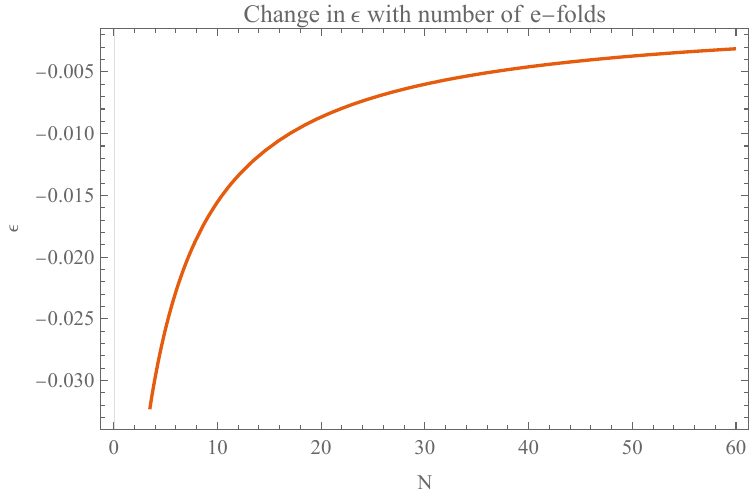}
     \caption{Slow roll parameter $\epsilon$ as a function of number of e-folds for $x=0.1$. The slow roll parameter is small and negative but increasing over time.}
    \label{fig:phantom}
\end{figure}
This result is interesting in the context of emerging observational evidence, such as DESI and Planck datasets, that hints at early-universe deviations from standard dark energy paradigms \cite{DESI:2025fii}. The phantom behavior of the solution may have implications for early dark energy. For the sake of being brief, we defer it to future works. Negative $\epsilon$ also indicates a violation of null-energy conditions, although not surprising due to the presence of non canonical kinetic terms in the field equation. \\

\section{Entropy of the early Universe} 
The action for gravity from entropy lends itself to an information theory interpretation similar to Boltzmann entropy. The Lagrangian for the theory
\begin{equation}
    \mathcal{L} = -Tr\log(\tilde{G}\tilde{g}^{-1}) = \log(W(r))
\end{equation}
where $W(r)$ ``counts" (In general, $W(r)$ defined here is real and not necessarily an integer), the number of degrees of freedom of geometry \cite{Bianconi:2024hts}. For a system containing $n$ macroscopic subsystems, the Boltzmann entropy is $S=\sum_{i=1}^n \log(W_{i}) $ where $W_i$ is the number of microscopic configurations of the subsystem. In a similar vein, for gravity from entropy, we have 
\begin{equation}
    S=\frac{1}{l_{pl}^4} \int \sqrt{-g} \log(W(r)) d^4x.
\end{equation}
Thus, the quantum relative entropy counts the number of
degrees of freedom of the metric. For the FLRW Universe, entropy is determined by the scale factor $a(t)$, and spatial volume $V$.
\begin{figure}[!ht]
    \centering
    \includegraphics[scale=0.53]{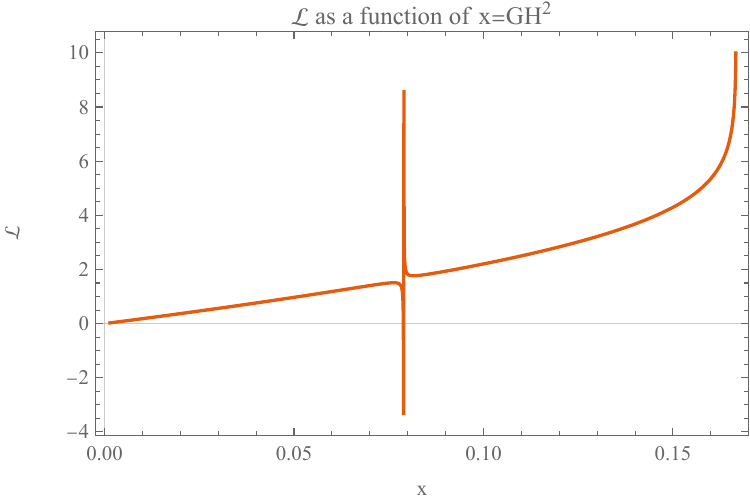}
     \caption{$\mathcal{L}$ as a function of $x=G H^2$ assuming 55 e-folds of inflation. Ignoring the peak near $x\sim0.08$ where our approximations are invalid $\mathcal{L}$ and consequently entropy is highest for slow roll solutions near $x\sim1/6$.}
    \label{fig:L}
\end{figure}
For solutions within the slow roll approximation discussed in the previous sections
\begin{equation}
    \begin{split}
         \mathcal{L} &= -Tr\log(\tilde{G}\tilde{g}^{-1}) \\
        &= \log 2 + \log 8
- 3 \log(1 - x)
- 3 \log\left(2 + x(-3 + \epsilon)\right)\\
&\;\;\;\;\;\;\;\;\;\;\;\;- \log\left(1 + 3x(-2 + \epsilon)\right)
- 3 \log\left(1 + x(-1 + \epsilon)\right)\\
&\;\;\;\;\;\;\;\;\;\;\;\;\;\;\;\;\;\;\;\;\;\;\;\;- \log\left(2 + 3x(-1 + \epsilon)\right).
    \end{split}
\end{equation}
For inflationary spacetime $a(t)\sim e^{H t}$, then
\begin{equation}
    S \simeq  V \frac{\mathcal{L}}{l_{pl}^4}  \int dt  e^{3H t} \simeq   \frac{V}{l_{pl}^3} \frac{\mathcal{L}}{ l_{pl} H} \left(e^{3H t}-1\right).
\end{equation}
Figure \ref{fig:L} depicts $\mathcal{L}$ as a function of $x=GH^2$. Ignoring $x \sim 0.08$ where our approximations break down,  $\mathcal{L}$ and the entropy are highest near $x\sim1/6$. The result in Figure (\ref{fig:L}) indicates that the inflationary solution with $\sqrt{\frac{1}{12\beta G}}$ has a higher entropy than the phantom-like solutions and slow-roll solutions where $H\ll $ $\sqrt{\frac{1}{12\beta G}}$. \\

 \section{Conclusions and discussions}
   This work develops and solves the modified Friedmann equations in the framework of entropic quantum gravity, where gravity emerges from the quantum relative entropy between the spacetime metric and a geometry–matter–induced metric referred in this paper as the gravity from entropy approach. The exponential expansion is inherent to the theory and does not require additional scalar fields or other exotic matter fields. The framework provides a UV-complete origin for the inflationary phase while requiring minimal parameters. Our analysis identified a high-entropy inflationary branch for $0.12 < x = 2 \beta G H^2 < 1/6$ that predicts $0.000142 \lesssim r \lesssim 0.012$ and $0.962 \lesssim n_s \lesssim 0.964$, consistent with current CMB constraints, and a phantom-like branch for $0.08 \lesssim x \lesssim 0.11$ with $w < -1$ and violation of the null energy condition, potentially relevant to early dark energy scenarios.
   
   The entropy interpretation shows that the inflationary branch corresponds to a higher number of geometric degrees of freedom than the phantom-like solutions. The violation of energy conditions hints at the possibility of bounce or cyclic cosmologies within this framework. A systematic study of scalar and tensor perturbations is needed to conclusively determine the predictions for CMB and GW (Gravitational Wave) observables. 
   Overall, the results suggest that both inflation and exotic early-universe behavior can emerge purely from the entropic gravity, offering a predictive and theoretically motivated alternative to scalar-field-driven inflation that warrants further investigation.\\
   
    \textit{\textbf{Acknowledgments}}-- We would like to acknowledge Prof. Ginestra Bianconi (Queen Mary University of London) for her insights and suggestions that helped to improve the draft significantly. The authors thank Prof. Ido Ben-Dayan (Ariel University, Israel) for his valuable comments and suggestions. The authors also gratefully acknowledge Dr. Utkarsh Kumar (University of Ottawa) and Dr. Vishnu Kakat (University of South Africa) for insightful discussions and helpful feedback. This work is supported in part by NSFC under Grant No. 11847239.

    \newpage
	
	\bibliography{ref.bib}
	
	\newpage
	\pagebreak

    \newpage
    \widetext
\begin{center}
\end{center}

\appendix
\section{Metric induced by geometry} \label{sec:appendix1}
Assume that metric induced by geometry is
\begin{equation}
     \tilde{\mathbf{G}} = \tilde{g}-\alpha \tilde{\mathbf{\mathcal{R}}} .
     \label{eq:gfiea}
\end{equation}
It is easy to see that
\begin{equation}
\begin{split}
    \tilde{\mathbf{G}}  \tilde{g}^{-1} = \mathbf{I}-\alpha \tilde{\mathbf{\mathcal{R}}}  \tilde{g}^{-1} = (1-\alpha R) \oplus \left( \delta_{\mu}^{\nu} -\alpha R_\mu^{\nu} \right)\; dx^{\mu} \otimes dx_{\nu} 
    \;\;\oplus \left(\frac{1}{2}
    \delta_{\mu\nu}^{\rho\sigma} - \alpha R_{\mu\nu}^{\;\;\;\rho \sigma}\right)  (dx^{\mu} \wedge dx^{\nu}) \otimes (dx_{\rho} \wedge dx_{\sigma}) 
    \end{split}
\end{equation}
where $ \delta_{\mu\nu}^{\rho\sigma}=\delta_{\mu}^{\rho}\delta_{\nu}^{\sigma}-\delta_{\mu}^{\sigma}\delta^{\rho}_{\nu}$. The elements of the flattened form of $ \frac{1}{2}
    \delta_{\mu\nu}^{\rho\sigma}-\alpha R_{\mu\nu}^{\;\;\;\rho \sigma} $ is
    \begin{equation}
     \left[  \frac{1}{2}
    \delta_{\mu\nu}^{\rho\sigma}-\alpha R_{\mu\nu}^{\;\;\;\rho \sigma} \right]_F= \left[\delta_{\mu\nu}^{\rho\sigma}-2\alpha R_{\mu\nu}^{\;\;\;\rho \sigma} \right].
    \end{equation}
    where $\mu<\nu$ and $\rho<\sigma$. 
According to the notation in \cite{Bianconi:2025rnd} for Schwarzschild black holes, we have $\alpha=\beta G$, and our choice of $\alpha=\frac{G}{2}$ corresponds to a choice of $\beta=\frac{1}{2}$.
	\onecolumngrid	
    \section{Modified Friedmann equations}   \label{sec:app2}
    By substituting the FLRW metric in equation \eqref{eq:modein} we obtain the modified Friedmann equations \eqref{eq:modfried}. Here we write the expanded version of the terms in the equation. \eqref{eq:modfried}. 
    The modified Ricci scalar for gravity from entropy is (indices here are not summed over)
  \begin{equation}
  \begin{split}
    R_{\mathcal{G}} &=  \mathrm{Tr} \left(g^{-1}\mathcal{G} \tilde{\mathcal{R}}\right) 
    =\frac{R}{1-G \frac{R}{2}} +\sum_{\mu}\frac{ R^{\mu}_{\mu}}{1-GR^{\mu}_{\mu}} +\sum^{\mu<\nu}_{\mu \nu}  \frac{2R_{\mu\nu}^{\;\;\;\mu \nu} }{\left(1- \frac{G}{2} R_{\mu\nu}^{\;\;\;\mu \nu}\right)}.
      \end{split}
  \end{equation}
    For FLRW metric 
\begin{equation}
\begin{aligned}
R_{\mathcal{G}} &=\frac{6\left(a(t)\ddot{a}(t) + \dot{a}(t)^{2} + k\right)}{a(t)^{2}}
\cdot \frac{1}{1 - \frac{3G\left(a(t)\ddot{a}(t) + \dot{a}(t)^{2} + k\right)}{a(t)^{2}}}
+ \frac{3\left(a(t)\ddot{a}(t) + 2\dot{a}(t)^{2} + 2k\right)}{a(t)^{2}}
\cdot \frac{1}{1 - \frac{G\left(a(t)\ddot{a}(t) + 2\dot{a}(t)^{2} + 2k\right)}{2a(t)^{2}}}
\\
&\quad
+ \frac{3\ddot{a}(t)}{a(t)} \cdot \frac{1}{1 - \frac{3G\ddot{a}(t)}{2a(t)}}
+ \frac{6\left(\dot{a}(t)^{2} + k\right)}{a(t)^{2}}
\cdot \frac{1}{1 - \frac{G\left(\dot{a}(t)^{2} + k\right)}{a(t)^{2}}}
+ \frac{6\ddot{a}(t)}{a(t)} \cdot \frac{1}{1 - \frac{G\ddot{a}(t)}{a(t)}}
\end{aligned}
\end{equation}

where $k$ is $\pm1$ or $0$ depending on the nature of spatial curvature of the Hubble Universe. 
spatial part of the Modified Ricci tensor $R_{11}$ is 

\begin{equation}
\begin{aligned}
&-\left(a(t)\ddot{a}(t) + 2\dot{a}(t)^{2} + 2k\right)\cdot
\frac{1}{k r^{2} - 1}\cdot
\frac{1}{1 - \frac{3G\left(a(t)\ddot{a}(t) + \dot{a}(t)^{2} + k\right)}{a(t)^{2}}}
+ \frac{2a(t)^{2}\left(a(t)\ddot{a}(t) + 2\dot{a}(t)^{2} + 2k\right)}
{a(t)\ddot{a}(t)G + 2\dot{a}(t)^{2}G + 2Gk - 2a(t)^{2}}
\cdot \frac{1}{k r^{2} - 1}
\\
&\quad
+ \frac{2a(t)^{2}\left(\dot{a}(t)^{2} + k\right)}
{\dot{a}(t)^{2}G - a(t)^{2} + Gk}\cdot \frac{1}{k r^{2} - 1}
- \frac{a(t)^{2}\ddot{a}(t)}{-G\ddot{a}(t) + a(t)}\cdot \frac{1}{k r^{2} - 1} = R^{\mathcal{G}}_{11}
\end{aligned}
\end{equation}
Temporal part of the modified Ricci tensor is given by 
\begin{equation}
\begin{aligned}
R^{\mathcal{G}}_{00}=-3 a(t)\ddot{a}(t)\cdot \frac{1}{-3 a(t)\ddot{a}(t) G - 3 \dot{a}(t)^{2} G + a(t)^{2} - 3 G k} - \frac{6 \ddot{a}(t)}{-3 G \ddot{a}(t) + 2 a(t)}
- \frac{3 \ddot{a}(t)}{- G \ddot{a}(t) + a(t)}
\end{aligned}
\end{equation}
$D_{\mu\nu}$ involves higher-order derivatives of the field $\tilde{\mathbf{\mathcal{G}}}$ and is diagonal. We split $D^{\mu \nu}$ in to different components and write expressions for the components below
\begin{equation}
  D^1_{\mu\nu}=  \left( \nabla^{\rho} \nabla_{\rho} g_{\mu \nu}
- \nabla_{\mu} \nabla_{\nu} \right) \mathcal{G}_{(0)} =0.
\end{equation}
Let 
\begin{equation}
  D^{2}_{\mu \nu}=   \nabla^{\rho} \nabla_{\nu} \left[ \mathcal{G}_{(1)} \right]_{(\rho \mu)} 
\end{equation}
then 
\begin{equation}
D^2_{11}=\frac{%
  \begin{aligned}
    12\,a(t)\Bigl(&
      a(t)\dot{a}(t)\bigl(\tfrac{a(t)\ddot{a}(t)G}{2}+\dot{a}(t)^2G+Gk - a(t)^2\bigr)\,\dddot{a}(t)\\
    &\quad +\,a(t)^2\,\ddot{a}(t)^3\,G
      -\,a(t)\bigl(-\tfrac{3\dot{a}(t)^2G}{2}+Gk+\tfrac{2a(t)^2}{3}\bigr)\,\ddot{a}(t)^2\\
    &\quad +\bigl(-4\dot{a}(t)^4G +(-4Gk-\tfrac{a(t)^2}{3})\dot{a}(t)^2 + \tfrac{2a(t)^2k}{3}\bigr)\,\ddot{a}(t)\\
    &\quad +\,2a(t)\dot{a}(t)^2\bigl(\dot{a}(t)^2 + k\bigr)
    \Bigr)\,G
  \end{aligned}
}{%
  \begin{aligned}
    (2\dot{a}(t)^2G &+ a(t)\ddot{a}(t)G - 2a(t)^2 + 2Gk)\\
    &\times (k\,r^2 - 1)\\
    &\times (3\,G\,\ddot{a}(t) - 2\,a(t))^2
  \end{aligned}
}
\end{equation}
and
\begin{equation}
D^2_{00}=\frac{%
  \begin{aligned}
  72\,\bigl(&
      a(t)\bigl(\tfrac{a(t)\,\ddot{a}(t)\,G}{2} + \dot{a}(t)^2\,G + G\,k - a(t)^2\bigr)^2
      \bigl(G\,\ddot{a}(t) - \tfrac{2\,a(t)}{3}\bigr)\,\dddot{a}(t)\\
    &\quad -\,2\,a(t)\bigl(\tfrac{a(t)\,\ddot{a}(t)\,G}{2} + \dot{a}(t)^2\,G + G\,k - a(t)^2\bigr)^2\,G\,\ddot{a}(t)^2\\
    &\quad +\,5\,\dot{a}(t)\Bigl(
        -\tfrac{G^2\,a(t)^2\,\ddot{a}(t)^3}{5}
        + G\,a(t)\bigl(\dot{a}(t)^2\,G + G\,k - \tfrac{2\,a(t)^2}{15}\bigr)\,\ddot{a}(t)^2\\
    &\qquad\quad
        + \bigl(G^2\,\dot{a}(t)^4 + (2\,G^2\,k - \tfrac{32\,G\,a(t)^2}{15})\,\dot{a}(t)^2
          + G^2\,k^2 - \tfrac{32\,G\,k\,a(t)^2}{15} + \tfrac{8\,a(t)^4}{15}\bigr)\,\ddot{a}(t)\\
    &\qquad\quad
        - \tfrac{2\,a(t)\,(\dot{a}(t)^2 + k)\,(\dot{a}(t)^2\,G + G\,k - 2\,a(t)^2)}{15}
      \Bigr)\,G\,\dddot{a}(t)\\
    &\quad -\,\tfrac{a(t)^2\,\ddot{a}(t)^5\,G^3}{4}
      -\,a(t)\bigl(10\,\dot{a}(t)^2\,G + G\,k - \tfrac{7\,a(t)^2}{6}\bigr)\,G^2\,\ddot{a}(t)^4\\
    &\quad -\,\bigl(
         -\tfrac{7\,G^2\,\dot{a}(t)^4}{2}
         +\bigl(-\tfrac{5\,G^2\,k}{2}-\tfrac{73\,G\,a(t)^2}{3}\bigr)\,\dot{a}(t)^2
         + G^2\,k^2 - \tfrac{8\,G\,k\,a(t)^2}{3} + \tfrac{5\,a(t)^4}{3}
       \bigr)\,G\,\ddot{a}(t)^3\\
    &\quad +\,\tfrac{2\,a(t)\bigl(
         -\tfrac{35\,G^2\,\dot{a}(t)^4}{2}
         +\bigl(-\tfrac{33\,G^2\,k}{2} - 28\,G\,a(t)^2\bigr)\,\dot{a}(t)^2
         + (G\,k - a(t)^2)^2
       \bigr)\,\ddot{a}(t)^2}{3}\\
    &\quad -\,\tfrac{10\,\dot{a}(t)^2\bigl(
         G^2\,\dot{a}(t)^4 + \bigl(2\,G^2\,k - \tfrac{21\,G\,a(t)^2}{5}\bigr)\,\dot{a}(t)^2
         + G^2\,k^2 - \tfrac{21\,G\,k\,a(t)^2}{5} - \tfrac{6\,a(t)^4}{5}
       \bigr)\,\ddot{a}(t)}{3}\\
    &\quad +\,\tfrac{4\,a(t)\,\dot{a}(t)^2\,(\dot{a}(t)^2 + k)\,(\dot{a}(t)^2\,G + G\,k - 3\,a(t)^2)}{3}
  \bigr)\,G
  \end{aligned}
}{%
  \begin{aligned}
    \bigl(2\,\dot{a}(t)^2\,G + a(t)\,\ddot{a}(t)\,G - 2\,a(t)^2 + 2\,G\,k\bigr)^2\\
    \times \bigl(3\,G\,\ddot{a}(t) - 2\,a(t)\bigr)^3
  \end{aligned}
}
\end{equation}
Similarly $D^3_{\mu\nu}=\frac{1}{2} \nabla^{\rho} \nabla_{\rho} 
\left[ \mathcal{G}_{(1)} \right]_{\mu \nu}$ and
\begin{equation}
\begin{aligned}
D^3_{11}=&\frac{%
  \begin{aligned}
    &-\,a(t)^{4}\,\ddot{a}(t)\,\ddddot{a}(t)\,G
    +2\,a(t)^{4}\,\dddot{a}(t)^{2}\,G
    -2\,a(t)^{3}\,\dot{a}(t)^{2}\,\ddddot{a}(t)\,G\\
    &\quad+10\,a(t)^{3}\,\dot{a}(t)\,\ddot{a}(t)\,\dddot{a}(t)\,G
    -3\,a(t)^{3}\,\ddot{a}(t)^{3}\,G
    -20\,a(t)^{2}\,\dot{a}(t)^{3}\,\dddot{a}(t)\,G\\
    &\quad+30\,a(t)^{2}\,\dot{a}(t)^{2}\,\ddot{a}(t)^{2}\,G
    -24\,a(t)\,\dot{a}(t)^{4}\,\ddot{a}(t)\,G
    +8\,\dot{a}(t)^{6}\,G\\
    &\quad+2\,a(t)^{5}\,\ddddot{a}(t)
    +4\,a(t)^{4}\,\dot{a}(t)\,\dddot{a}(t)
    +6\,a(t)^{4}\,\ddot{a}(t)^{2}
    -36\,a(t)^{3}\,\dot{a}(t)^{2}\,\ddot{a}(t)\\
    &\quad-2\,a(t)^{3}\,\ddddot{a}(t)\,G\,k
    +24\,a(t)^{2}\,\dot{a}(t)^{4}
    -20\,a(t)^{2}\,\dot{a}(t)\,\dddot{a}(t)\,G\,k\\
    &\quad-2\,a(t)^{2}\,\ddot{a}(t)^{2}\,G\,k
    -16\,a(t)\,\dot{a}(t)^{2}\,\ddot{a}(t)\,G\,k
    +16\,\dot{a}(t)^{4}\,G\,k\\
    &\quad-8\,a(t)^{3}\,\ddot{a}(t)\,k
    +24\,a(t)^{2}\,\dot{a}(t)^{2}\,k
    +8\,a(t)\,\ddot{a}(t)\,G\,k^{2}
    +8\,\dot{a}(t)^{2}\,G\,k^{2}
  \end{aligned}
}%
{%
  \bigl(k\,r^{2}-1\bigr)\,
  \bigl(-a(t)\,\ddot{a}(t)\,G-2\,\dot{a}(t)^{2}\,G+2\,a(t)^{2}-2\,G\,k\bigr)^{3}
}
\\[1em]
&\quad
+\,\frac{%
  2\,\bigl(\dddot{a}(t)\,a(t)^{2}
        -4\,\dot{a}(t)^{3}
        +3\,\dot{a}(t)\,\ddot{a}(t)\,a(t)
        -4\,\dot{a}(t)\,k\bigr)\,G\,a(t)^{2}\,\dot{a}(t)
}%
{%
  \bigl(k\,r^{2}-1\bigr)\,
  \bigl(-a(t)\,\ddot{a}(t)\,G-2\,\dot{a}(t)^{2}\,G+2\,a(t)^{2}-2\,G\,k\bigr)^{2}
}
\\[1em]
&\quad
+\,\frac{%
  \begin{aligned}
    &\bigl(13\,\dot{a}(t)\,\ddot{a}(t)^{2}\,a(t)^{2}\,G
      -8\,\dot{a}(t)^{3}\,\ddot{a}(t)\,a(t)\,G
      +3\,\dddot{a}(t)\,\ddot{a}(t)\,a(t)^{3}\,G\\
    &\quad-8\,\dot{a}(t)^{5}\,G
      -14\,\dot{a}(t)\,\ddot{a}(t)\,a(t)^{3}
      -8\,\dot{a}(t)\,\ddot{a}(t)\,a(t)\,G\,k
      +16\,\dot{a}(t)^{3}\,a(t)^{2}\\
    &\quad-16\,\dot{a}(t)^{3}\,G\,k
      -2\,\dddot{a}(t)\,a(t)^{4}
      +16\,\dot{a}(t)\,a(t)^{2}\,k
      -8\,\dot{a}(t)\,G\,k^{2}\bigr)\,G\,\dot{a}(t)\,a(t)
  \end{aligned}
}%
{%
  \bigl(k\,r^{2}-1\bigr)\,
  \bigl(3\,G\,\ddot{a}(t)-2\,a(t)\bigr)\,
  \bigl(-a(t)\,\ddot{a}(t)\,G-2\,\dot{a}(t)^{2}\,G+2\,a(t)^{2}-2\,G\,k\bigr)^{2}
}
\end{aligned}
\end{equation}
\begin{equation}
\begin{aligned}
&\quad  D_{00}^3 =
\,9\,G\,\dot a(t)\,\dddot a(t)\,\frac{1}{\bigl(-3\,G\,\ddot a(t)+2\,a(t)\bigr)^2}
\;-\;6\,\dot a(t)^2\,\frac{1}{\bigl(-3\,G\,\ddot a(t)+2\,a(t)\bigr)^2}  \\
&-3G\,
\frac{%
  3\,\ddot a(t)^3\,G+
  (-3G\,a(t)\,\ddot a(t) +2\,a(t)^2 
  -4\,a(t)\,\dot a(t)\,\dddot a(t) ) \ddddot a(t)
  -6\,\ddot a(t)\dot a(t)\dddot a(t)\,G
  +6\,a(t)\dddot a(t)^2\,G
  -2\,a(t)\ddot a(t)^2
  +4\,\ddot a(t)\,\dot a(t)^2
}%
{(3\,G\,\ddot a(t)-2\,a(t))^3}
\\
&\quad
-\frac{3\,\dot a(t)^2}{\,a(t)\,\bigl(-3\,G\,\ddot a(t)+2\,a(t)\bigr)\,}
\;+\;\frac{6\,\dot a(t)^2}
      {-\,a(t)\,\ddot a(t)\,G \;-\;2\,\dot a(t)^2\,G \;+\;2\,a(t)^2 \;-\;2\,G\,k}
\end{aligned}
\end{equation}
We also define $D^4_{\mu\nu}= \frac{1}{2} \nabla^{\eta} \nabla_{\eta} 
\left[ \mathcal{G}_{(1)} \right]_{\rho \nu} g_{\mu \nu}$, then  
\begin{equation}
\begin{split}
D^4_{11}=\frac{1}{%
  (k\,r^{2}-1)\,
  \bigl(2\dot a(t)^{2}G + a(t)\ddot a(t)\,G - 2a(t)^{2} + 2Gk\bigr)^{2}\,
  \bigl(3G\ddot a(t) - 2a(t)\bigr)^{3}
}
\\[0.5em]
\times
\bigl(
  -\,a(t)^{4}\,\ddot a(t)\,\ddddot a(t)\,G
  +2\,a(t)^{4}\,\dddot a(t)^{2}\,G
  -2\,a(t)^{3}\,\dot a(t)^{2}\,\ddddot a(t)\,G
  +10\,a(t)^{3}\,\dot a(t)\,\ddot a(t)\,\dddot a(t)\,G\\
\quad
  -3\,a(t)^{3}\,\ddot a(t)^{3}\,G
  -20\,a(t)^{2}\,\dot a(t)^{3}\,\dddot a(t)\,G
  +30\,a(t)^{2}\,\dot a(t)^{2}\,\ddot a(t)^{2}\,G
  -24\,a(t)\,\dot a(t)^{4}\,\ddot a(t)\,G\\
\quad
  +8\,\dot a(t)^{6}\,G
  +2\,a(t)^{5}\,\ddddot a(t)
  +4\,a(t)^{4}\,\dot a(t)\,\dddot a(t)
  +6\,a(t)^{4}\,\ddot a(t)^{2}
  -36\,a(t)^{3}\,\dot a(t)^{2}\,\ddot a(t)\\
\quad
  -2\,a(t)^{3}\,\ddddot a(t)\,G\,k
  +24\,a(t)^{2}\,\dot a(t)^{4}
  -20\,a(t)^{2}\,\dot a(t)\,\dddot a(t)\,G\,k
  -2\,a(t)^{2}\,\ddot a(t)^{2}\,G\,k\\
\quad
  -16\,a(t)\,\dot a(t)^{2}\,\ddot a(t)\,G\,k
  +16\,\dot a(t)^{4}\,G\,k
  -8\,a(t)^{3}\,\ddot a(t)\,k
  +24\,a(t)^{2}\,\dot a(t)^{2}\,k\\
\quad
  +8\,a(t)\,\ddot a(t)\,G\,k^{2}
  +8\,\dot a(t)^{2}\,G\,k^{2}
\bigr)
\end{split}
\end{equation}
and
\begin{equation}
\begin{split}
D^4_{00}=&\frac{1}{%
  a(t)\,(3G\,\ddot a(t)-2\,a(t))^{3}\,
  \bigl(2\,\dot a(t)^{2}G + a(t)\,\ddot a(t)G - 2\,a(t)^{2} + 2\,G\,k\bigr)^{2}
}
\\[0.5em]
&\quad\times 36\,\Bigl(
  a(t)^{2}\bigl(\tfrac{a(t)\,\ddot a(t)\,G}{2}+\dot a(t)^{2}G+Gk-a(t)^{2}\bigr)^{2}
    \bigl(G\,\ddot a(t)-\tfrac{2\,a(t)}{3}\bigr)\,\ddddot a(t)\\
&\qquad
  -2\,a(t)^{2}\bigl(\tfrac{a(t)\,\ddot a(t)\,G}{2}+\dot a(t)^{2}G+Gk-a(t)^{2}\bigr)^{2}
    G\,\dddot a(t)^{2}\\
&\qquad
  +8\,\dot a(t)\,a(t)\,\Bigl(
      -\tfrac{G^{3}a(t)^{2}\,\ddot a(t)^{3}}{32}
      +G^{2}a(t)\bigl(\dot a(t)^{2}G+Gk-\tfrac{25\,a(t)^{2}}{48}\bigr)\,\ddot a(t)^{2}\\
&\qquad\quad
      +\bigl(G^{2}\dot a(t)^{4}+(2\,G^{2}k-\tfrac{7\,G\,a(t)^{2}}{3})\dot a(t)^{2}
      +G^{2}k^{2}-\tfrac{7\,Gk\,a(t)^{2}}{3}+\tfrac{23\,a(t)^{4}}{24}\bigr)\,G\,\ddot a(t)\\
&\qquad\quad
      -\tfrac{\bigl(\dot a(t)^{2}G+Gk-\tfrac{3\,a(t)^{2}}{2}\bigr)\,a(t)\,
        \bigl(\dot a(t)^{2}G+Gk-\tfrac{a(t)^{2}}{2}\bigr)}{3}
    \Bigr)\,\dddot a(t)\\
&\qquad
  +\frac{5\,\ddot a(t)^{5}\,a(t)^{3}\,G^{3}}{4}
  +\frac{a(t)^{2}\bigl(-\tfrac{19\,\dot a(t)^{2}G}{2}+Gk-\tfrac{23\,a(t)^{2}}{3}\bigr)\,G^{2}\,\ddot a(t)^{4}}{2}\\
&\qquad
  -4\,\Bigl(
    -\tfrac{G^{2}\,\dot a(t)^{4}}{2}
    +\bigl(\tfrac{G^{2}k}{2}-\tfrac{83\,G\,a(t)^{2}}{24}\bigr)\,\dot a(t)^{2}
    +G^{2}k^{2}-\tfrac{11\,Gk\,a(t)^{2}}{12}-\tfrac{3\,a(t)^{4}}{4}
  \Bigr)\,a(t)\,G\,\ddot a(t)^{3}\\
&\qquad
  +\Bigl(
    -12\,\dot a(t)^{6}G^{3}
    +\bigl(\tfrac{10\,G^{2}a(t)^{2}}{3}-24\,G^{3}k\bigr)\,\dot a(t)^{4}
    +\bigl(-12\,G^{3}k^{2}+8\,G^{2}k\,a(t)^{2}-13\,G\,a(t)^{4}\bigr)\,\dot a(t)^{2}\\
&\qquad\quad
    +\tfrac{14\,a(t)^{2}\bigl(G^{2}k^{2}-Gk\,a(t)^{2}-\tfrac{a(t)^{4}}{7}\bigr)}{3}
  \Bigr)\,\ddot a(t)^{2}\\
&\qquad
  +\frac{32\,a(t)\,\bigl(
    \dot a(t)^{6}G^{2}
    +\bigl(2\,G^{2}k-\tfrac{G\,a(t)^{2}}{8}\bigr)\,\dot a(t)^{4}
    +\bigl(G^{2}k^{2}-\tfrac{Gk\,a(t)^{2}}{4}+\tfrac{5\,a(t)^{4}}{16}\bigr)\,\dot a(t)^{2}
    -\tfrac{a(t)^{2}k(Gk-a(t)^{2})}{8}
  \bigr)\,\ddot a(t)}{3}\\
&\qquad
  -\frac{8\,G\,a(t)^{2}\,\dot a(t)^{2}\,(\dot a(t)^{2}+k)^{2}}{3}
\Bigr)\,,
\end{split}
\end{equation}
Now $D^5_{\mu\nu}= \nabla^{\eta} \nabla^{\nu} 
\left[ \mathcal{G}_{(2)} \right]_{\mu \rho \eta}$ and $D^6_{\mu\nu}=\nabla^{\eta} \nabla^{\nu} 
\left[ \mathcal{G}_{(2)} \right]_{\eta \mu \nu} $, for FLRW spacetime
\begin{equation}
\begin{split}
D^5_{11}=&\frac{1}{%
  2\,(k\,r^{2}-1)\,
  \bigl(\dot a(t)^{2}G - a(t)^{2} + G\,k\bigr)^{2}\,
  \bigl(G\,\ddot a(t) - a(t)\bigr)^{3}
}\\
&\times a(t)\Bigl(
  a(t)^{2}\bigl(\dot a(t)^{2}G - a(t)^{2} + G\,k\bigr)^{2}
    \bigl(G\,\ddot a(t)-a(t)\bigr)\,\ddddot a(t)\\
&\quad
  -2\,G\,a(t)^{2}\bigl(\dot a(t)^{2}G - a(t)^{2} + G\,k\bigr)^{2}
    \dddot a(t)^{2}\\
&\quad
  +6\,a(t)\bigl(G\,\ddot a(t)-\tfrac{a(t)}{3}\bigr)\,
    \bigl(\dot a(t)^{2}G - a(t)^{2} + G\,k\bigr)^{2}\,
    \dot a(t)\,\dddot a(t)\\
&\quad
  +2\,G^{2}\,a(t)^{2}\bigl(-\dot a(t)^{2}G + G\,k - a(t)^{2}\bigr)\,
    \ddot a(t)^{4}\\
&\quad
  -3\,a(t)\bigl(-G^{2}\dot a(t)^{4}-\tfrac{10\,G\,a(t)^{2}\,\dot a(t)^{2}}{3}
       +G^{2}k^{2}-a(t)^{4}\bigr)\,G\,\ddot a(t)^{3}\\
&\quad
  +\bigl(
    -6\,\dot a(t)^{6}G^{3}
    +\bigl(-12\,G^{3}k - G^{2}a(t)^{2}\bigr)\dot a(t)^{4}\\
&\qquad\quad
    +\bigl(-6\,G^{3}k^{2}+4\,G^{2}k\,a(t)^{2}-16\,G\,a(t)^{4}\bigr)\dot a(t)^{2}
    +5\,a(t)^{2}G^{2}k^{2}-4\,G\,k\,a(t)^{4}-a(t)^{6}
  \bigr)\,\ddot a(t)^{2}\\
&\quad
  +6\,a(t)\bigl(
    \dot a(t)^{6}G^{2}
    +\bigl(2\,G^{2}k+\tfrac{2\,G\,a(t)^{2}}{3}\bigr)\dot a(t)^{4}\\
&\qquad\quad
    +\bigl(G^{2}k^{2}+\tfrac{G\,k\,a(t)^{2}}{3}+a(t)^{4}\bigr)\dot a(t)^{2}
    -\tfrac{a(t)^{2}k(G\,k-a(t)^{2})}{3}
  \bigr)\,\ddot a(t)\\
&\quad
  -2\,a(t)^{2}\,\dot a(t)^{2}\,(\dot a(t)^{2}+k)\,(\dot a(t)^{2}G+G\,k+a(t)^{2})
\Bigr)\,G
\end{split}
\end{equation}
and
\begin{equation}
\begin{split}
D^5_{00}=&\frac{1}{%
  2\,a(t)\,
  \bigl(\dot a(t)^2\,G \;-\; a(t)^2 \;+\; G\,k \bigr)\,
  \bigl(-\,G\,\ddot a(t) \;+\; a(t)\bigr)^2
}\;
\times\;3\,\dot a(t)\,G\\
&\quad\times
\Bigl(
  -\,3\,\ddot a(t)\,\dot a(t)^3\,G
  +\,a(t)\,\dot a(t)^2\,\dddot a(t)\,G
  +\,2\,\ddot a(t)^2\,a(t)\,\dot a(t)\,G\\
&\quad\qquad
  +\,2\,a(t)\,\dot a(t)^3
  -\,\ddot a(t)\,a(t)^2\,\dot a(t)
  -\,3\,\ddot a(t)\,\dot a(t)\,G\,k\\
&\quad\qquad
  -\,a(t)^3\,\dddot a(t)
  +\,a(t)\,\dddot a(t)\,G\,k
  +\,2\,a(t)\,\dot a(t)\,k
\Bigr)\,.  
\end{split}
\end{equation}
Also 
\begin{equation}
    D^7_{\mu\nu}=\frac{1}{2} \left[ \nabla^{\rho}, \nabla^{\eta} \right] 
\left[ \mathcal{G}_{(2)} \right]_{\rho \eta \mu \nu}=0.
\end{equation}
Now 
\begin{equation}
    \begin{split}
        D_{\mu\nu}=-D^2_{\mu\nu}+D^3_{\mu\nu}+D^4_{\mu\nu}+D^5_{\mu\nu}+D^6_{\mu\nu}.
    \end{split}
\end{equation}
The cosmological constant term $\Lambda_{\mathcal{G}}$ is 
\begin{equation}
\begin{split}
\Lambda_{\mathcal{G}}=\frac{1}{2G}\Bigl(
&\frac{1}{1 - \frac{3G\bigl(a(t)\ddot a(t) + \dot a(t)^{2} + k\bigr)}{a(t)^{2}}}
- 11
- \ln\!\Bigl(\frac{1}{1 - \frac{3G\bigl(a(t)\ddot a(t) + \dot a(t)^{2} + k\bigr)}{a(t)^{2}}}\Bigr)\\
&\quad
+ \frac{6\,a(t)^{2}}{-\,a(t)\ddot a(t)\,G - 2\,\dot a(t)^{2}\,G + 2\,a(t)^{2} - 2\,Gk}
+ \frac{2\,a(t)}{-3G\ddot a(t) + 2\,a(t)}\\
&\quad
- 3\,\ln\!\Bigl(\frac{2\,a(t)^{2}}{-\,a(t)\ddot a(t)\,G - 2\,\dot a(t)^{2}\,G + 2\,a(t)^{2} - 2\,Gk}\Bigr)
- \ln\!\Bigl(\frac{2\,a(t)}{-3G\ddot a(t) + 2\,a(t)}\Bigr)\\
&\quad
+ \frac{3\,a(t)^{2}}{-\,\dot a(t)^{2}\,G + a(t)^{2} - Gk}
+ \frac{3\,a(t)}{-\,G\ddot a(t) + a(t)}\\
&\quad
- 3\,\ln\!\Bigl(\frac{a(t)^{2}}{-\,\dot a(t)^{2}\,G + a(t)^{2} - Gk}\Bigr)
- 3\,\ln\!\Bigl(\frac{a(t)}{-\,G\ddot a(t) + a(t)}\Bigr)
\Bigr)
\end{split}
\end{equation}
\subsection{Modified Friedmann equations for $k=0$ in terms of Hubble parameter} \label{sec:app2a}
In this section, we rewrite the modified equations of motion in terms of the Hubble parameter for the special case of a flat Universe. If $H$ is the Hubble parameter, we have the following.
\begin{equation}
R^{\mathcal{G}}=\frac{a(t)^{2}\Bigl(
  -16
  - \frac{6}{-1 + G\,H(t)^{2}}
  - \frac{12}{-2 + 3\,G\,H(t)^{2} + G\,\dot H(t)}
  - \frac{3}{-1 + G\bigl(H(t)^{2} + \dot H(t)\bigr)}
  + \frac{1 + 3\,G\,H(t)^{2}}{1 - 3\,G\bigl(2\,H(t)^{2} + \dot H(t)\bigr)}
\Bigr)}{3\,G}
\end{equation}
\begin{equation}
R^{\mathcal{G}}_{00}=\frac{1}{G} \left[ 6 + \frac{1 - 3 G H^2}{-1 + 6 G H^2 + 3 G \dot{H}} + \frac{3}{-1 + G\left(H^2 + \dot{H}\right)} + \frac{4}{-2 + 3 G\left(H^2 + \dot{H}\right)} \right].
\end{equation}
\begin{equation}
R^{\mathcal{G}}_{11}=\frac{a^2}{3 G} \left[ -16 + \frac{6}{1 - G H^2} + \frac{12}{2 - 3 G H^2 - G \dot{H}} + \frac{3}{1 - G\left(H^2 + \dot{H}\right)} + \frac{1 + 3 G H^2}{1 - 3 G\left(2 H^2 + \dot{H}\right)} \right].
\end{equation}
\begin{equation}
\begin{aligned}
D^{2}_{1 1}=& \frac{2 G a^2}{\left(-2 + 3 G H^2 + G \dot{H}\right)\left(2 - 3 G (H^2 + \dot{H})\right)^2}
\\
& \quad \times \Big(
-2 \dot{H}\,\big(21 G H^4 + \dot{H}(-2 + 3 G \dot{H}) + 2 H^2(-7 + 9 G \dot{H})\big)
- 3 H\,\big(-2 + 3 G H^2 + G \dot{H}\big)\,\ddot{H}
\Big)
\end{aligned}
\end{equation}
\begin{equation}
\begin{aligned}
D^2_{00}&= \frac{3 G}{\big(-2 + 3 G H^2 + G \dot{H}\big)\,\big(-2 + 3 G (H^2 + \dot{H})\big)^3}
\\
& \quad \times \Big(
18 G^2 H^6 \dot{H}
+ 45 G^2 H^5 \ddot{H}
- 12 G H^3 (5 + G \dot{H})\,\ddot{H}
- H \big(-2 + G \dot{H}\big)\big(10 + 9 G \dot{H}\big)\,\ddot{H}
\\
& \qquad
- 3 G H^4 \big( 2 \dot{H} (4 + 9 G \dot{H}) - 3 G \dddot{H} \big)
- 2 H^2 \big( -4 \dot{H} - 12 G \dot{H}^2 + 9 G^2 \dot{H}^3 + 9 G^2 \ddot{H}^2
- 6 G (-1 + G \dot{H})  \big)
\\
&\qquad
+ \big(-2 + G \dot{H}\big) \big( -4 \dot{H}^2 + 6 G \dot{H}^3 + 3 G \dot{H} \dddot{H}
- 2 \big( 3 G \ddot{H}^2 + \dddot{H} \big) \big)
\Big).
\end{aligned}
\end{equation}
\begin{equation}
\begin{aligned}
D^3_{11}&= -\,\frac{G\,a^2}{
\big(-2 + 3 G H^2 + G \dot{H}\big)^{3}\,
\big(-2 + 3 G (H^2 + \dot{H})\big)
}
\\
& \quad \times \Big(
198 G^{2} H^{6} \dot{H}
+ 81 G^{2} H^{5} \ddot{H}
+ 36 G H^{3} \big(-3 + G \dot{H}\big)\,\ddot{H}
- 3 H \big(-2 + 3 G \dot{H}\big)\big(6 + 5 G \dot{H}\big)\,\ddot{H}
\\
& \qquad\quad
+ 3 G H^{4} \big(-88 \dot{H} + 26 G \dot{H}^{2} + 3 G \dddot{H}\big)
+ \big(-2 + 3 G \dot{H}\big)\!\left( 6 \dot{H}^{2} \big(-2 + G \dot{H}\big)
- 2 G \ddot{H}^{2}
+ \big(-2 + G \dot{H}\big) \dddot{H} \right)
\\
& \qquad\quad
- 2 H^{2} \left(
-44 \dot{H}
+ 44 G \dot{H}^{2}
+ 43 G^{2} \dot{H}^{3}
+ 3 G^{2} \ddot{H}^{2}
- 6 G \big(-1 + G \dot{H}\big) \dddot{H}
\right)
\Big),
\end{aligned}
\end{equation}

\begin{equation}
\begin{aligned}
D_{00}^3&= \frac{3 G}{\big(-2 + 3 G H^2 + G \dot{H}\big)\,\big(-2 + 3 G (H^2 + \dot{H})\big)^3}
\\
& \quad \times \Big(
18 G^2 H^6 \dot{H}
+ 45 G^2 H^5 \ddot{H}
- 12 G H^3 (5 + G \dot{H})\,\ddot{H}
- H \big(-2 + G \dot{H}\big)\big(10 + 9 G \dot{H}\big)\,\ddot{H}
\\
& \qquad
- 3 G H^4 \big( 2 \dot{H} (4 + 9 G \dot{H}) - 3 G \dddot{H} \big)
- 2 H^2 \big( -4 \dot{H} - 12 G \dot{H}^2 + 9 G^2 \dot{H}^3 + 9 G^2 \ddot{H}^2
- 6 G (-1 + G \dot{H}) \dddot{H} \big)
\\
& \qquad
+ \big(-2 + G \dot{H}\big) \big( -4 \dot{H}^2 + 6 G \dot{H}^3 + 3 G \dot{H} \dddot{H}
- 2 \big( 3 G \ddot{H}^2 + \dddot{H} \big) \big)
\Big).
\end{aligned}
\end{equation}
\begin{equation}
\begin{aligned}
D^4_{11}&= -\,\frac{3 G a^{2}}{\left(-2 + 3 G H^{2} + G \dot{H}\right)^{2}
\left(-2 + 3 G \left(H^{2} + \dot{H}\right)\right)^{3}}
\\
& \quad \times \Big(
540 G^{3} H^{8} \dot{H}
+ 8 \dot{H}^{2} \left(-2 + G \dot{H}\right) \left(-1 + G \dot{H}\right) \left(-2 + 3 G \dot{H}\right)
\\
& \qquad
+ 108 G^{2} H^{6} \dot{H} \left(-10 + 7 G \dot{H}\right)
+ 12 G H^{4} \dot{H} \left(60 + G \dot{H} \left(-90 + 31 G \dot{H}\right)\right)
\\
& \qquad
+ 4 H^{2} \dot{H} \left(-40 + G \dot{H} \left(108 + 7 G \dot{H} \left(-14 + 3 G \dot{H}\right)\right)\right)
+ 243 G^{3} H^{7} \ddot{H}
\\
& \qquad
+ 9 G^{2} H^{5} \left(-54 + 17 G \dot{H}\right) \ddot{H}
- 3 G H^{3} \left(-108 + G \dot{H} \left(68 + 5 G \dot{H}\right)\right) \ddot{H}
\\
& \qquad
- H \left( 2 + G \dot{H} \right)
\left( 36 + G \dot{H} \left( -52 + 21 G \dot{H} \right) \right) \ddot{H}
\\
& \qquad
- 54 G^{3} H^{4} \ddot{H}^{2}
- 36 G^{2} H^{2} \left(-2 + G \dot{H}\right) \ddot{H}^{2}
- 6 G \left(-2 + G \dot{H}\right)^{2} \ddot{H}^{2}
\\
& \qquad
+ \left(-2 + 3 G H^{2} + G \dot{H}\right)^{2}
\left(-2 + 3 G \left(H^{2} + \dot{H}\right)\right) \dddot{H}
\Big)
\end{aligned}
\end{equation}
\begin{equation}
\begin{gathered}
D^4_{00}=\frac{1}{
\left(-2 + 3 G H^{2} + G \dot{H}\right)^{2}
\left(-2 + 3 G \left(H^{2} + \dot{H}\right)\right)^{3}
}
\\
\times
3 G \Big[
324 G^3 H^{8} \dot{H}
+ 324 G^{2} H^{6} \dot{H} \left(-2 + G \dot{H}\right)
+ 8 \dot{H}^{2} \left(-2 + G \dot{H}\right) \left(-1 + G \dot{H}\right) \left(-2 + 3 G \dot{H}\right)
\\
\quad
+ 36 G H^{4} \dot{H} \left(12 + G \dot{H} \left(-14 + 3 G \dot{H}\right)\right)
+ 12 H^{2} \dot{H} \left(-8 + G \dot{H} \left(20 + 3 G \dot{H} \left(-6 + G \dot{H}\right)\right)\right)
\\
\quad
+ 189 G^{3} H^{7} \ddot{H}
+ 63 G^{2} H^{5} \left(-6 + G \dot{H}\right) \ddot{H}
- 3 G H^{3} \left(-84 + G \dot{H} \left(28 + 19 G \dot{H}\right)\right) \ddot{H}
\\
\quad
+ H \left(-56 + G \dot{H} \left(28 + G \dot{H} \left(38 - 27 G \dot{H}\right)\right)\right) \ddot{H}
- 54 G^{3} H^{4} \ddot{H}^{2}
- 36 G^{2} H^{2} \left(-2 + G \dot{H}\right) \ddot{H}^{2}
\\
\quad
- 6 G \left(-2 + G \dot{H}\right)^{2} \ddot{H}^{2}
+ \left(-2 + 3 G H^{2} + G \dot{H}\right)^{2}
\left(-2 + 3 G \left(H^{2} + \dot{H}\right)\right) \dddot{H}
\Big].
\end{gathered}
\end{equation}
Now $D^5_{\mu\nu}= \nabla^{\eta} \nabla^{\nu} 
\left[ \mathcal{G}_{(2)} \right]_{\mu \rho \eta}$ and $D^6_{\mu\nu}=\nabla^{\eta} \nabla^{\nu} 
\left[ \mathcal{G}_{(2)} \right]_{\eta \mu \nu} $, for FLRW spacetime
\begin{equation}
\begin{gathered}
D^5_{11}=D^6_{11}=\frac{1}{
2 \left(-1 + G H^{2}\right)^{2}
  \left(-1 + G \left(H^{2} + \dot{H}\right)\right)^{3}
}
\\
\times\;
G\, a^{2} \Big[
-8 G^{3} H^{8} \dot{H}
+ 4 \dot{H}^{2}
- 6 G \dot{H}^{3}
+ 2 G^{2} \dot{H}^{4}
- 6 G^{3} H^{7} \ddot{H}
+ 2 H \left(3 + G \dot{H}\right) \ddot{H}
\\
\quad
+ 2 G^{2} H^{5} \left(9 + G \dot{H}\right) \ddot{H}
- 2 G H^{3} \left(9 + 2 G \dot{H}\right) \ddot{H}
+ 2 G \ddot{H}^{2}
+ \dddot{H}
- G \dot{H}\, \dddot{H}
\\
\quad
+ G^{2} H^{6} \left(24 \dot{H} - G \dddot{H}\right)
+ G H^{4} \Big( 2 \dot{H} \big(-12 + G \dot{H} (2 + G \dot{H})\big)
                 + 2 G^{2} \ddot{H}^{2}
                 + G (3 - G \dot{H}) \dddot{H} \Big)
\\
\quad
+ H^{2} \Big( 2 \dot{H} \big(4 + G \dot{H} \big(-4 + G \dot{H} (2 + G \dot{H})\big)\big)
              - 4 G^{2} \ddot{H}^{2}
              + G (-3 + 2 G \dot{H}) \dddot{H} \Big)
\Big]
\end{gathered}
\end{equation}
and
\begin{equation}
\begin{gathered}
D^5_{00}=D^6_{00}=\frac{1}{
2 \left(-1 + G H^{2}\right)
  \left(-1 + G \left(H^{2} + \dot{H}\right)\right)^{2}
}
\\
\times\;
3 G H \Big(
2 H \dot{H} \left(-2 + 2 G H^{2} + G \dot{H}\right)
+ \left(-1 + G H^{2}\right) \ddot{H}
\Big).
\end{gathered}
\end{equation}

\begin{equation}
\begin{gathered}
\Lambda_{\mathcal{G}}=\frac{1}{2G}
\times
\Big(
-\Big[
11
+ \frac{3}{-1 + G H^{2}}
+ 3 \log\!\left(\frac{1}{1 - G H^{2}}\right)
+ 3 \log\!\left(-\frac{2}{-2 + 3 G H^{2} + G \dot{H}}\right)
+ 3 \log\!\left(\frac{1}{1 - G \left(H^{2} + \dot{H}\right)}\right)
\\[4pt]
\quad
+ \log\!\left(-\frac{2}{-2 + 3 G \left(H^{2} + \dot{H}\right)}\right)
+ \log\!\left(\frac{1}{1 - 3 G \left(2 H^{2} + \dot{H}\right)}\right)
+ \frac{6}{-2 + 3 G H^{2} + G \dot{H}}
+ \frac{1}{-1 + 6 G H^{2} + 3 G \dot{H}}
\\[4pt]
\quad
+ \frac{3}{-1 + G \left(H^{2} + \dot{H}\right)}
+ \frac{2}{-2 + 3 G \left(H^{2} + \dot{H}\right)}
\Big]
\Big).
\end{gathered}
\end{equation}
\section{Derivation of Inflationary solution} \label{sec:app3}

From the time component of the modified Friedmann equation \eqref{eq:modfried} we have
\begin{equation}
\begin{aligned}
H''(t)=& \left[ -1 + G H(t)^{2} + G \dot{H}(t) \right]^{2}
  \left[ -2 + 3 G H(t)^{2} + G \dot{H}(t) \right]^{2}
  \left[ -2 + 3 G H(t)^{2} + 3 G \dot{H}(t) \right]^{2} \\
& \quad \times \Bigg\{
    \frac{3 H(t)^{2}}{-1 + G H(t)^{2}}
    + \frac{3 \left[ H(t)^{2} + \dot{H}(t) \right]}{-1 + G H(t)^{2} + G \dot{H}(t)}
    + \frac{3 \left[ 3 H(t)^{2} + \dot{H}(t) \right]}{-2 + 3 G H(t)^{2} + G \dot{H}(t)} \\
& \qquad
    - \frac{1458 G^{6} H(t)^{12} \dot{H}(t)}{\left[ -1 + G H(t)^{2} \right]
      \left[ -1 + G H(t)^{2} + G \dot{H}(t) \right]^{2}
      \left[ -2 + 3 G H(t)^{2} + G \dot{H}(t) \right]^{2}
      \left[ -2 + 3 G H(t)^{2} + 3 G \dot{H}(t) \right]^{2}} \\
& \qquad
    + \frac{3 \left[ H(t)^{2} + \dot{H}(t) \right]}{-2 + 3 G H(t)^{2} + 3 G \dot{H}(t)}
    + \frac{3 \left[ 2 H(t)^{2} + \dot{H}(t) \right]}{-1 + 6 G H(t)^{2} + 3 G \dot{H}(t)} \\
& \qquad
    - \frac{54 G^{5} H(t)^{10} \dot{H}(t) \left[ -105 + 88 G \dot{H}(t) \right]}{\Den}\\
    &\qquad + \frac{6 G \dot{H}(t)^{2} \left[ -1 + G \dot{H}(t) \right]^{2} \left[ 4 - 8 G \dot{H}(t) + 3 G^{2} \dot{H}(t)^{2} \right]}{\Den} \\
& \qquad
    - \frac{18 G^{4} H(t)^{8} \dot{H}(t) \left[ 489 - 829 G \dot{H}(t) + 326 G^{2} \dot{H}(t)^{2} \right]}{\Den} \\& \qquad
    - \frac{6 G^{3} H(t)^{6} \dot{H}(t) \left[ -1137 + 2929 G \dot{H}(t) - 2342 G^{2} \dot{H}(t)^{2} + 564 G^{3} \dot{H}(t)^{3} \right]}{\Den} \\
& \qquad
    - \frac{6 G^{2} H(t)^{4} \dot{H}(t) \left[ 440 - 1541 G \dot{H}(t) + 1891 G^{2} \dot{H}(t)^{2} - 934 G^{3} \dot{H}(t)^{3} + 147 G^{4} \dot{H}(t)^{4} \right]}{\Den} \\
& \qquad
    - \frac{6 G H(t)^{2} \dot{H}(t) \left[ -68 + 312 G \dot{H}(t) - 535 G^{2} \dot{H}(t)^{2} + 415 G^{3} \dot{H}(t)^{3} - 137 G^{4} \dot{H}(t)^{4} + 12 G^{5} \dot{H}(t)^{5} \right]}{\Den} \\
& \qquad
    - \frac{3 \left[ H(t)^{2} + \dot{H}(t) \right] \left[ 6 + 33 G^{2} H(t)^{4} - 22 G \dot{H}(t) + 18 G^{2} \dot{H}(t)^{2} + 17 G H(t)^{2} \left( -2 + 3 G \dot{H}(t) \right) \right]}{\left[ -1 + G H(t)^{2} + G \dot{H}(t) \right] \left[ -2 + 3 G H(t)^{2} + 3 G \dot{H}(t) \right] \left[ -1 + 6 G H(t)^{2} + 3 G \dot{H}(t) \right]} \\
& \qquad
    - \frac{1}{2G} \Bigg[ 11 + \frac{3}{-1 + G H(t)^{2}} + 3 \ln\!\left( \frac{1}{1 - G H(t)^{2}} \right)
      + \ln\!\left( \frac{1}{1 - 6 G H(t)^{2} - 3 G \dot{H}(t)} \right) \\
& \qquad\qquad
      + 3 \ln\!\left( -\frac{1}{-1 + G H(t)^{2} + G \dot{H}(t)} \right)
      + 3 \ln\!\left( -\frac{2}{-2 + 3 G H(t)^{2} + G \dot{H}(t)} \right)
      + \ln\!\left( -\frac{2}{-2 + 3 G H(t)^{2} + 3 G \dot{H}(t)} \right) \\
& \qquad
      + \frac{3}{-1 + G H(t)^{2} + G \dot{H}(t)}
      + \frac{6}{-2 + 3 G H(t)^{2} + G \dot{H}(t)}
      + \frac{2}{-2 + 3 G H(t)^{2} + 3 G \dot{H}(t)}
      + \frac{1}{-1 + 6 G H(t)^{2} + 3 G \dot{H}(t)}
    \Bigg]
\Bigg\} \\
& \quad \bigg/ \Bigg[ 3 G H(t) \Big( 32 + 117 G^{4} H(t)^{8} - 120 G \dot{H}(t) + 164 G^{2} \dot{H}(t)^{2} - 96 G^{3} \dot{H}(t)^{3} + 21 G^{4} \dot{H}(t)^{4} \\
& \qquad\qquad
    + 12 G^{3} H(t)^{6} \left( -28 + 27 G \dot{H}(t) \right)
    + 2 G^{2} H(t)^{4} \left( 182 - 348 G \dot{H}(t) + 159 G^{2} \dot{H}(t)^{2} \right) \\
& \qquad\qquad
    + 4 G H(t)^{2} \left( -44 + 125 G \dot{H}(t) - 114 G^{2} \dot{H}(t)^{2} + 33 G^{3} \dot{H}(t)^{3} \right)
\Big) \Bigg]
\end{aligned}
\label{eq:apph2}
\end{equation}
Substituting the expression \eqref{eq:apph2}  in the second modified Friedmann equation (rewritten in number of efolds) and assuming the slow roll scenario where $H \sim e^{-\epsilon N}$ and expanding up to first order in $\epsilon$, we obtain an expression for $\epsilon$ given by
\begin{equation}
\begin{aligned}
\epsilon & = 
1/
\Bigg[(1 - 6x)\, (2 - 5x + 3x^{2})^{2}\,
\Big( x \big( 240 - 1872x + 6108x^{2} - 10656x^{3} + 10488x^{4} - 5532x^{5} + 1225x^{6} \big) \\&
\qquad \qquad+ 2 N \big( 288 - 2160x + 6804x^{2} - 11512x^{3} + 11040x^{4} - 5700x^{5} + 1241x^{6} \big) \Big)\,
\Big( \ln\!\big( \tfrac{1}{1 - 6x} \big) \Big)^{2} \Bigg]
\\[0.25em]
& \qquad \times
\Big(
(2 - 3x)^{2} (1 - x)^{2} (1 - 6x) \big( 8 - 20x + 13x^{2} \big)
\Big[ \ln\!\Big( \tfrac{1}{1 - 6x} \Big) + 4 \ln\!\Big( \tfrac{2}{2 - 3x} \Big) + 6 \ln\!\Big( \tfrac{1}{1 - x} \Big) \Big]
\\
& \qquad\quad \times
\Big( 9x \big( 80 - 392x + 730x^{2} - 612x^{3} + 195x^{4} \big)
+ \big( -36 + 222x - 552x^{2} + 692x^{3} - 437x^{4} + 111x^{5} \big) \ln\!\Big( \tfrac{1}{1 - 6x} \Big)
\\
&\qquad\qquad
+ 4 \big( -36 + 222x - 552x^{2} + 692x^{3} - 437x^{4} + 111x^{5} \big) \ln\!\Big( \tfrac{2}{2 - 3x} \Big) \\ &\qquad \qquad 
+ 6 \big( -36 + 222x - 552x^{2} + 692x^{3} - 437x^{4} + 111x^{5} \big) \ln\!\Big( \tfrac{1}{1 - x} \Big)
\Big)
\Big)
\\[0.5em]
& \qquad
+ 16 (1 - 6x) (2 - 5x + 3x^{2})^{2}
\Big( x \big( 240 - 1872x + 6108x^{2} - 10656x^{3} + 10488x^{4} - 5532x^{5} + 1225x^{6} \big) \\& \qquad \qquad
+ 2 N \big( 288 - 2160x + 6804x^{2} - 11512x^{3} + 11040x^{4} - 5700x^{5} + 1241x^{6} \big) \Big)
\Big( \ln\!\big( \tfrac{2}{2 - 3x} \big) \Big)^{2}
\\[0.5em]
& \qquad
- 2 (2 - 5x + 3x^{2}) \ln\!\Big( \tfrac{1}{1 - 6x} \Big)
\Big\{
3x (8 - 20x + 13x^{2})
\Big( x \big( -368 + 4544x - 20044x^{2} + 44056x^{3} - 52554x^{4} + 32757x^{5} - 8406x^{6} \big)
\\
& \qquad\qquad\qquad\qquad
+ 4 N \big( -216 + 2568x - 12180x^{2} + 30717x^{3} - 45122x^{4} + 38927x^{5} - 18357x^{6} + 3663x^{7} \big) \Big)
\\
& \qquad\qquad
- 4 \big( -2 + 17x - 33x^{2} + 18x^{3} \big)
\Big( x \big( 240 - 1872x + 6108x^{2} - 10656x^{3} + 10488x^{4} - 5532x^{5} + 1225x^{6} \big)\\& \qquad \qquad
+ 2 N \big( 288 - 2160x + 6804x^{2} - 11512x^{3} + 11040x^{4} - 5700x^{5} + 1241x^{6} \big) \Big)
\ln\!\Big( \tfrac{2}{2 - 3x} \Big)
\\
& \qquad\qquad
- 6 \big( -2 + 17x - 33x^{2} + 18x^{3} \big)
\Big( x \big( 240 - 1872x + 6108x^{2} - 10656x^{3} + 10488x^{4} - 5532x^{5} + 1225x^{6} \big)\\ & \qquad \qquad
+ 2 N \big( 288 - 2160x + 6804x^{2} - 11512x^{3} + 11040x^{4} - 5700x^{5} + 1241x^{6} \big) \Big)
\ln\!\Big( \tfrac{1}{1 - x} \Big)
\Big\}
\\[0.5em]
& \qquad
- 24 (2 - 5x + 3x^{2}) \ln\!\Big( \tfrac{2}{2 - 3x} \Big)
\Big\{
x (8 - 20x + 13x^{2})
\Big( x \big( -368 + 4544x - 20044x^{2} + 44056x^{3} - 52554x^{4} + 32757x^{5} - 8406x^{6} \big)
\\
& \qquad\qquad\qquad\qquad
+ 4 N \big( -216 + 2568x - 12180x^{2} + 30717x^{3} - 45122x^{4} + 38927x^{5} - 18357x^{6} + 3663x^{7} \big) \Big)
\\
& \qquad\qquad
- 2 \big( -2 + 17x - 33x^{2} + 18x^{3} \big)
\Big( x \big( 240 - 1872x + 6108x^{2} - 10656x^{3} + 10488x^{4} - 5532x^{5} + 1225x^{6} \big)\\& \qquad \qquad
+ 2 N \big( 288 - 2160x + 6804x^{2} - 11512x^{3} + 11040x^{4} - 5700x^{5} + 1241x^{6} \big) \Big)
\ln\!\Big( \tfrac{1}{1 - x} \Big)
\Big\}
\\[0.5em]
& \qquad
- 9 \Big[
x^{2} (8 - 20x + 13x^{2})^{2}
\Big( -384 + 5772x - 34984x^{2} + 99736x^{3} - 145712x^{4} + 106641x^{5} - 31194x^{6}
\\
& \qquad\qquad\qquad\qquad
+ 12 N \big( 120 - 1268x + 5072x^{2} - 10236x^{3} + 11208x^{4} - 6381x^{5} + 1485x^{6} \big) \Big)
\\
& \qquad\qquad
+ 4x (16 - 80x + 150x^{2} - 125x^{3} + 39x^{4})
\Big( x \big( -368 + 4544x - 20044x^{2} + 44056x^{3} - 52554x^{4} + 32757x^{5} - 8406x^{6} \big)
\\
& \qquad\qquad\qquad\qquad
+ 4 N \big( -216 + 2568x - 12180x^{2} + 30717x^{3} - 45122x^{4} + 38927x^{5} - 18357x^{6} + 3663x^{7} \big) \Big)
\ln\!\Big( \tfrac{1}{1 - x} \Big)
\\
& \qquad\qquad
- 4 (1 - 6x) (2 - 5x + 3x^{2})^{2}
\Big( x \big( 240 - 1872x + 6108x^{2} - 10656x^{3} + 10488x^{4} - 5532x^{5} + 1225x^{6} \big)\\ & \qquad \qquad
+ 2 N \big( 288 - 2160x + 6804x^{2} - 11512x^{3} + 11040x^{4} - 5700x^{5} + 1241x^{6} \big) \Big)
\Big( \ln\!\big( \tfrac{1}{1 - x} \big) \Big)^{2}
\Big]
\Bigg]
\end{aligned}
\label{eq:fep}
\end{equation}
where $x= G H^2$. Expanding equation \eqref{eq:fep} for small x we have
\begin{equation}
\epsilon \sim \frac{3}{2 \left( -8 + 3N \right)}
- \frac{(11 + 3N)\,x}{4 \left( -8 + 3N \right)^{2}}
\end{equation}

\end{document}